\documentclass[12pt]{article}
\usepackage{mathrsfs}
\usepackage{amssymb}
\usepackage{amsmath}
\usepackage[amsmath,thmmarks]{ntheorem}
\usepackage{cite}

\usepackage{graphicx}
\usepackage{dcolumn}
\usepackage{bm}
\usepackage{slashed}

\newcommand{\be}{\begin{eqnarray}}
\newcommand{\ee}{\end{eqnarray}}
\newcommand{\ra}{\rightarrow}

\begin{document}

\begin{titlepage}

\vskip.40cm
\begin{center}
{\Large {\bf Reconstructing a $Z^{\prime}$ Lagrangian using the LHC and low-energy data}} \vskip.5cm
\end{center}
\vskip0.2cm

\begin{center}
{\bf Ye Li, Frank Petriello and Seth Quackenbush}
\end{center}
\vskip 8pt
\begin{center}
{\it Department of Physics, University of Wisconsin, Madison, WI 53706, USA} \\
\end{center}

\vglue 0.3truecm

\begin{abstract}
\vskip 3pt \noindent

We study the potential of the LHC and future low-energy experiments to precisely measure the underlying model parameters of a new $Z^{\prime}$ boson.  We emphasize the complimentary information 
obtained from both on- and off-peak LHC dilepton data, from the future Q-weak measurement of the weak charge of the proton, and from a proposed measurement of parity violation in low-energy 
M\o{}ller scattering.  We demonstrate the importance of off-peak LHC data and Q-weak for removing sign degeneracies between $Z^{\prime}$ couplings that occur if only on-peak LHC data is studied.  A future precision measurement of low-energy M\o{}ller scattering can resolve a scaling degeneracy between quark and lepton couplings that remains after analyzing LHC dilepton data, permitting an extraction of the individual $Z^{\prime}$ couplings rather than combinations of them.  We study how precisely $Z^{\prime}$ properties can be extracted for LHC integrated luminosities ranging from 
a few inverse femtobarns to super-LHC values of an inverse attobarn.  For the several example cases studied with $M_{Z'}=1.5$ TeV, we find that coupling combinations can be determined with relative uncertainties reaching $\pm 30\%$ with 30 ${\rm fb}^{-1}$ of integrated luminosity, while $\pm 50\%$ is possible with 10 ${\rm fb}^{-1}$.  With SLHC luminosities of 1 ab$^{-1}$, we find that products of quark and lepton couplings can be probed to $\pm 10\%$.

\end{abstract}

\end{titlepage}
\newpage

\section{Introduction \label{intro}}

$Z'$ gauge bosons arise in most constructions of physics beyond the Standard Model (SM).   They appear in grand unified theories such as $SO(10)$~\cite{mohapatra} and $E(6)$~\cite{Hewett:1988xc}, in Little Higgs models~\cite{Schmaltz:2005ky}, and in theories with extra space-time dimensions~\cite{Hewett:2002hv}.  They often appear as messengers 
which connect the SM to hidden sectors, such as in Hidden Valley models~\cite{Strassler:2006im}.  Due to their pervasiveness in models of new physics, 
significant effort has been devoted to 
searching for them experimentally.  $Z'$ states that decay to lepton pairs have a simple, clean experimental signature and can easily be searched for at 
high-energy colliders.  Current direct search limits from the Tevatron require the $Z'$ mass to be greater than about 1 TeV when its couplings to SM fermions are identical to those of the $Z$ boson~\cite{Aaltonen:2008ah}.

Since they are so prevalent in models of new physics, and it worthwhile to study what could be learned from a $Z'$ discovery at the LHC.  This also serves as a useful 
LHC benchmark for how well new-physics Lagrangian parameters can be experimentally determined.  Since the experimental signature is clean and the QCD uncertainties have been studied and found to be fairly small~\cite{Fuks:2007gk,Petriello:2008zr,Coriano:2008wf}, it is likely that the couplings of a discovered $Z'$ can be studied with reasonable 
accuracy to probe the high scale theory that gave rise to it.  Many studies of how to discover, identify, or measure $Z'$ properties and couplings to SM particles have 
been performed~\cite{zprevs,delAguila:1993ym,Cvetic:1995zs,Dittmar:2003ir,Rizzo:2003ug,Carena:2004xs,Coriano:2008wf,Rizzo:2009pu}.  Recently, we performed a detailed analysis using a fully differential next-to-leading order QCD simulation to quantify how well all possible $Z'$ couplings could be extracted from 
on-peak data at the LHC assuming realistic QCD, parton distribution function (PDF), and statistical errors~\cite{Petriello:2008zr}.  Our study found that four combinations that probe 
both the parity-symmetric and parity-violating couplings of the $Z'$ could be probed, with some combinations measurable with roughly 15\% precision assuming $100 \, {\rm fb}^{-1}$ of data.  

This previous study contains several limitations.  More statistical power is possible if data from below the $Z'$ peak is utilized.  Also, additional differentiation between various $Z'$ couplings is 
possible with off-peak data.  On-peak, the only combinations of couplings that appears 
are $(q \times e)^2$, where $q$ and $e$ denote arbitrary quark and lepton couplings.  Two degeneracies exist in this expression: the relative signs between various $q\times e$ combinations are not determined, and the scaling $q \to yq, e\to e/y$, with $y$ an arbitrary constant, leaves the expression unchanged.  The first can be removed by Drell-Yan production of the $Z'$ off peak at the LHC, where the cross section contains a linear dependence on $q\times e$.  The removal of the second scaling degeneracy requires 
observation of the $Z'$ in a channel other than Drell-Yan at the LHC, or in a different experiment.

In this manuscript we extend the previous study to address several of these issues for the example case of a $Z'$ with mass $M_{Z'} = 1.5$ TeV.  We perform a detailed simulation of the $Z'$ signal and SM Drell-Yan background to study the potential of the LHC for a precision extraction of the model parameters.  We include errors arising from statistics, uncertainties in PDFs, and missing higher-order QCD and electroweak corrections.  We study how well 
couplings can be measured assuming different 
energy and luminosity scenarios at the LHC, including collisions at both $\sqrt{s}=10$ and 14 TeV, and integrated luminosities ranging from 3 ${\rm fb}^{-1}$ to super-LHC amounts of 1 ${\rm ab}^{-1}$.  If a $Z'$ is found at the LHC, additional information on its parameters can be obtained from various low-energy measurements.  Much of this information is complementary, as many low-energy observables depend on different coupling combinations than LHC measurements.  We study several planned future experiments with the potential for observing and studying a deviation from their SM 
expectations, and that may help remove the degeneracies described above.  We focus on the Q-weak experiment~\cite{VanOers:2007zz}, which plans to measure with high precision the weak charge of the proton in electron-proton scattering, and a proposed Jefferson Laboratory measurement of the effective weak-mixing angle through parity violation in low-energy M\o{}ller scattering~\cite{mollerup}.   This second experiment is of special interest.  Since it depends only upon the leptonic $Z'$ couplings, it can potentially break the scaling degeneracy between quark and lepton couplings 
to the $Z'$.  Our primary findings are summarized below. 

\begin{itemize}

\item Discrimination between the studied test models at the 68\% and 90\% confidence level with as little as 3 ${\rm fb}^{-1}$ of LHC data at the center-of-mass energy $\sqrt{s}=10$ TeV.  With 
10 ${\rm fb}^{-1}$ of integrated luminosity at this energy, measurements of the $c_q$ parameters defined in Refs.~\cite{Carena:2004xs,Petriello:2008zr} with relative uncertainties of 
$\pm 50\%$ are possible, while $\pm 30\%$ errors become possible with 30 ${\rm fb}^{-1}$.

\item With high luminosities of 300 ${\rm fb}^{-1}$ at $\sqrt{s}=14$ TeV, the relative signs between the various $q\times e$ combinations are determined, and measurements of these combinations 
approaching $10\%$ precision are possible.  

\item An interesting feature we derive from our analysis is an independent extraction of the width $\Gamma_{Z'}$ to a few GeV precision obtained without fitting the $Z'$ Breit-Wigner peak.  Since the on-peak cross section scales as $\sim 1 / \Gamma_{Z'}$ while the off-peak result contains no width dependence, comparison of data from different invariant mass bins allows a good determination.

\item A precision measurement of the weak charge of the proton with Q-weak helps remove sign degeneracies between the $q\times e$ that remain after LHC running.

\item With the inclusion of a future measurement of the effective weak-mixing angle in M\o{}ller scattering, the individual $q$ and $e$ charges can be separately determined.  We demonstrate 
for an example model how a combination of LHC Drell-Yan data and M\o{}ller scattering make this possible.  LHC measurements restrict the $Z'$ model parameters to lie in a wedge in the $e_L$ versus $e_R$ coupling plane, while M\o{}ller scattering restrict the couplings to a hyperbolic region.  A combination of the two experiments localizes the couplings to a point in this plane, within errors.

\end{itemize}

Our paper is organized as follows.  In Section~\ref{sec:theory} we review the parameters which describe $Z'$ production at the LHC and introduce the various test models we use to illustrate 
our analysis procedure.  We describe the LHC observables which we use to analyze $Z'$ properties in Section~\ref{sec:LHCmeas}, and discuss the expected low-energy measurements of 
the proton weak charge and the effective weak-mixing angle in Section~\ref{sec:lowmeas}.  We present our analysis of $Z'$ couplings at the LHC and in the Q-weak experiment, which 
are both sensitive to the $q \times e$ coupling combinations, in Section~\ref{sec:LHCanalysis}.  We study both low and high integrated luminosities in this Section, detailing the 
expected uncertainties in the parameter extractions for a variety of different running scenarios.  A summary showing how well all $q \times e$ couplings can be measured as a function of integrated luminosity is given in Section~\ref{sec:parex}.  In Section~\ref{sec:moller} we consider the constraints imposed by a proposed 
M\o{}ller scattering experiment, and demonstrate how a precision measurement at J-Lab allows the determination of the individual $q$ and $e$ charges.  Finally, we conclude in 
Section~\ref{sec:conc}.

\section{Theoretical Framework}
\label{sec:theory}

In this section we motivate a parametrization of $Z'$ models which can be probed by future experiments, while remaining as model-neutral as possible.  We also describe examples of common models which fit into our parametrization, and which we use to test our procedure for extracting these parameters.

\subsection{$Z'$ parameters \label{params}}

At the LHC, the smoking-gun signal for a $Z'$ is a new resonance in dileptons.  Other particles that decay into such final states, such as Kaluza-Klein gravitons, 
can be distinguished from a $Z'$ by study of the lepton angular distributions~\cite{Allanach:2000nr,Osland:2009tn}. The production of this resonance depends on the following parameters:
\begin{itemize}
\item the location of the resonance, i.e. the mass of the state, $M_{Z'}$;
\item the width of the resonance, $\Gamma_{Z'}$;
\item the couplings of the quarks which produce the $Z'$, $u_L^i, u_R^i, d_L^i, d_R^i$ with the index $i$ denoting the generation\footnotemark ;
\item the couplings of the charged leptons in the observed decay mode, $e_L^i, e_R^i$, and the neutrino coupling $v_L^i$.
\end{itemize}
\footnotetext{We have absorbed any overall coupling into these parameters to simplify the discussion and analysis.}
While the $Z'$ may couple to and decay into any number of other particles, from $W$ bosons to new exotic fermions, these parameters only enter into the Drell-Yan channel we consider through the width.

One cannot hope to determine all of these parameters in a completely model-independent way.  To proceed, we must make some assumptions on the parameter space.  Fortunately, 
several restrictions are very well motivated by experimental data.  We first make the assumption that the $Z'$ couplings are generation independent.  If they were not, a $Z'$ light enough to be discovered at the LHC would generate large flavor changing neutral currents in contradiction with experiment.  $Z'$ bosons with coupling strengths 
approximately the same as the SM $Z$ and ${\cal O}(1)$ coupling differences between the first two generations must have masses greater than roughly $10^3$ TeV due to 
constraints on the generated $\Delta F=2$ four-fermion operators~\cite{Bona:2007vi}.  Constraints on $Z'$ bosons with different couplings to third-generation fermions are weaker; deviations from 
our parameterization for top quarks, bottom quarks, or tau leptons can be searched for explicitly by observation of $Z'$ decay into these final states~\cite{Barger:2006hm,Godfrey:2008vf,Anderson:1992jz}.  Second, we assign states in a given $SU(2)_L$ doublet the same coupling to the $Z'$.  If the generator of the new gauge group to which the $Z'$ belongs does not commute with $SU(2)_L$, the $Z'$ would generically couple to the electroweak symmetry-breaking sector, inducing $Z-Z'$ mixing.  Specific models which would protect against this mixing, such as Kaluza-Klein $Z'$s, should be distinguishable from the class of models considered here~\cite{Rizzo:2003ug}.  LEP $Z$-pole measurements 
restrict the mixing angle to be smaller than a few thousandths of a radian~\cite{Abreu:1995, Erler:2009jh}.  We take this magnitude of mixing to be negligible.  We note that arranging the Z-Z' mixing angle to be this small requires, for the models considered, either fine-tuning the charges and vevs in the Higgs sector which gives mass to the Z and Z' bosons, or setting these parameters smaller than their expected values by roughly an order of magnitude~\cite{Carena:2004xs}.  There are many models in the literature which avoid these issues, but the $E_6$ models we use as examples remain a useful benchmark, particularly for comparing with other studies.  These restrictions leave the following 
seven $Z'$ parameters to be determined:
\begin{itemize}
\item the mass $M_{Z'}$ and the width $\Gamma_{Z'}$;
\item $e_L$ and $e_R$, the couplings to the lepton doublet and right-handed electron;
\item $q_L$, the quark doublet coupling;
\item $u_R$ and $d_R$, the right-handed quark couplings.
\end{itemize}

\subsection{Test models \label{models}}

To illustrate how well our analysis strategy determines $Z'$ properties, we test it on several example models.  Several well-known grand unified theories (GUTs) fit into the framework described above.  We take three such examples to illustrate how well their parameters could be determined if they are found.  Two of these arise from the exceptional group $E_6$, and one from $SO(10)$.  Below we briefly describe these models and list their couplings.  For a more comprehensive introduction to these specific examples, we refer the reader to several excellent reviews~\cite{mohapatra,Hewett:1988xc}.   Our procedure is by no means limited by these choices; we choose these common examples simply to demonstrate the efficacy of our model-independent approach.  This method can be used for any $Z'$ effective theory satisfying the coupling assumptions outlined above.

\begin{itemize}
\item \underline{$E_6$ Models}: $E_{6}$ models are described by the breaking chain
\be
E_{6} \ra SO(10) \times U(1)_{\psi} \ra SU(5) \times U(1)_{\chi} \times U(1)_{\psi} \ra SM \times U(1)_{\beta}
\ee
where
\be
Z' = Z'_{\chi} \cos\beta + Z'_{\psi} \sin\beta
\ee
is the lightest new boson arising from this breaking.  In this paper we take the $\chi$ model ($\beta = 0$), and the $\psi$ model ($\beta = \pi/2$), as representatives, though the whole family fits into our framework.

\item \underline{Left-right models}: We also consider a left-right model coming from the symmetry group $SU(2)_{R} \times SU(2)_{L} \times U(1)_{B-L}$.  Left-right models can arise from the following breaking of $SO(10)$:
\be
SO(10) \ra SU(3) \times SU(2)_{L} \times SU(2)_{R} \times U(1)_{B-L} .
\ee
The $Z'$ in left-right models couples to the current
\be
J^{mu}_{LR} = \alpha_{LR} J^{mu}_{3R} - 1/2\alpha_{LR} J^{mu}_{B-L}, 
\ee
with $\alpha_{LR} = \sqrt{(c^{2}_{W}g^{2}_{R}/s^{2}_{W}g^{2}_{L}) - 1}$, and $g_{L} = e/\cos\theta_{W}$.  We examine the symmetric case $g_{L} = g_{R}$, where $\alpha_{LR} \simeq 1.59$ if one takes the on-shell value of $\sin^2\theta_W$.
\end{itemize}

The fermionic couplings of these models are summarized in Table~\ref{couplings}.  For the standard definition of the $E_6$ models, we take the overall coupling to have its GUT-scale relation to the EM coupling down to the $Z'$ scale.  An overall $e/\cos\theta_W$ has been factored out.  We will later also consider a version of the $\chi$ model with a slightly larger overall coupling to demonstrate what is gained from potential measurements in low-energy experiments.  The overall factor of $e/\cos\theta_W \simeq 0.36$ in this model has been replaced by $1/2$.  We denote this model $\chi^*$.

\begin{table}[ht]
\begin{center}
\begin{tabular}{| c || c | c | c |}
\hline
 & $\chi$ & $\psi$ & $LR$ \\ \hline \hline
 $q_{L}$ & $\frac{-1}{2\sqrt{6}}$ & $\frac{\sqrt{10}}{12}$ & $\frac{-1}{6\alpha_{LR}}$ \\ \hline
 $u_{R}$ & $\frac{1}{2\sqrt{6}}$ & $\frac{-\sqrt{10}}{12}$ & $\frac{-1}{6\alpha_{LR}}+\frac{\alpha_{LR}}{2}$ \\ \hline
 $d_{R}$ & $\frac{-3}{2\sqrt{6}}$ & $\frac{-\sqrt{10}}{12}$ & $\frac{-1}{6\alpha_{LR}}-\frac{\alpha_{LR}}{2}$ \\ \hline
 $e_{L}$ & $\frac{3}{2\sqrt{6}}$ & $\frac{\sqrt{10}}{12}$ & $\frac{1}{2\alpha_{LR}}$ \\ \hline
 $e_{R}$ & $\frac{1}{2\sqrt{6}}$ & $\frac{-\sqrt{10}}{12}$ & $\frac{1}{2\alpha_{LR}}-\frac{\alpha_{LR}}{2}$ \\ \hline
\end{tabular} 
\end{center}
\caption{Fermion couplings to the $Z'$ for the considered models.  An overall $e/\cos\theta_{W}$ has been factored out. \label{couplings}}
\end{table}

\section{LHC Analysis Framework}
\label{sec:LHCmeas}

The primary means by which we will probe $Z'$ properties will be through examining the dilepton mode at the LHC.  We describe here the structure of the $Z'$ 
cross section, present the observables we utilize for various choices of integrated luminosity, and discuss the details of our simulation procedure.

\subsection{LHC measurements}
\label{sec:bins}

We would like to determine seven parameters: $M_{Z'}, \Gamma_{Z'}, q_L, u_R, d_R, e_L,$ and $e_R$.  The location of the resonance peak in the invariant mass of the dileptons should determine $M_Z'$ extremely well with a sufficient number of events; we take the error in this measurement to be negligible.  For this analysis, we take as an example $M_{Z'} = 1.5$ TeV.  For most models, much lighter $Z'$s are excluded by experiment~\cite{Abreu:1995, Aaltonen:2008ah}.  For significantly heavier $Z'$s, testing shows that there is not much improvement over our previous on-peak analysis,~\cite{Petriello:2008zr}.  In principle, the width can be determined from the resonance shape if the experimental resolution in invariant mass does not dominate the natural width of the $Z'$.  With enough statistics, widths smaller than the experimental 
resolution could possibly be probed by fitting the observed spectrum to a convolution of $Z'$ resonance shape with the experimental Gaussian resolution.  How well this can be done requires a detailed detector-level study and is beyond the scope of this analysis.  We make no assumption on how well the width can be determined, beyond the requirement that one be able to define an ``on-peak'' bin containing most of the resonance, to be explained below.  For an experimental study of both the $Z'$ mass and width measurement at the LHC, see Ref.~\cite{atlas_sim}.

One can reconstruct the following kinematic variables from the dilepton measurements at the LHC.  Taking the four-momenta of the electron and positron to be $k_1$ and $k_2$, we have, with $k_{Z'} = k_1 + k_2$:

\begin{itemize}
\item the invariant mass of the pair, $M^2 = k_{Z'}^2$;
\item the reconstructed rapidity of the $Z'$, $Y=\frac{1}{2} \log\frac{k_{Z'}^0 + k_{Z'}^3}{k_{Z'}^0 - k_{Z'}^3}$;
\item the scattering angle of the electron, $\theta^*$, measured in the Collins-Soper frame \cite{Collins:1977iv}.  The sign of $\cos\theta^*$ is chosen to be positive if $k_1^3$ and $k_{Z'}^3$ have the same sign, i.e., the electron is scattered in the same direction as the $Z'$ is boosted down the beam pipe.
\end{itemize}
The differential cross section of $pp \ra e^+e^-$ can then be written in the following form:
\begin{align}
\frac{d^2\sigma}{dYd\cos\theta^*dM^2} = \sum_{q=u,d} & [a_1^q(M_{Z'}, \Gamma_{Z'})(q_R^2+q_L^2)(e_R^2+e_L^2) + a_2^q(M_{Z'}, \Gamma_{Z'})(q_R^2-q_L^2)(e_R^2-e_L^2) \notag \\
+ & b_1^q(M_{Z'})q_Re_L + b_2^q(M_{Z'})q_Re_R] \notag \\
+ & b_3(M_{Z'})q_Le_L + b_4(M_{Z'})q_Le_R + c . \label{crs-sect}
\end{align}
The coefficients $a, b,$ and $c$ contain all kinematic and PDF dependence.  The $c$ term is the contribution from the SM $\gamma$ and $Z$.  To predict a measurement, one only has to integrate them over the appropriate region, and then input the dependence on the $Z'$ couplings.  By choosing the regions to integrate over (measurement bins) appropriately, one can solve the system of equations and gain coupling information.  We now motivate the choice of bins.

\begin{itemize}
 \item \underline{Forwards/Backwards}: At the parton level, terms symmetric in the quark-lepton scattering angle go as $(q_R^iq_R^j+q_L^iq_L^j)(e_R^ie_R^j+e_L^ie_L^j)$, where $i,j \in \gamma, Z, Z'$.  Terms antisymmetric in the scattering angle go as $(q_R^iq_R^j-q_L^iq_L^j)(e_R^ie_R^j-e_L^ie_L^j)$.  We could separate these terms by binning events depending on whether they are forwards or backwards scattered; $F+B$ is symmetric and $F-B$ is antisymmetric.  We do not have direct access to the partons, however, and the LHC is a pp collider, so there is no preferred direction for the quarks.  We define ``forwards'' to be $F=\int_0^1 d\cos\theta^*$ and ``backwards'' to be $B=\int_{-1}^0 d\cos\theta^*$.  The advantage of defining the direction in terms of the $Z'$ boost is that this direction correlates with the quark direction, and therefore one still gets useful information from the asymmetry even though the LHC is a $pp$ collider~\cite{Langacker:1984dc}.

\item \underline{Rapidity}: The higher the rapidity $Z'$, the more likely it is to have come from a valence $u$ quark and sea $\bar{u}$ quark.  This is due to the dominance of valence $u$ at high $x$.  In principle, to separate $u$ from $d$-type quarks, one needs only two rapidity bins, where the dividing line is chosen such that the bins contain approximately the same number of events~\cite{delAguila:1993ym}.  There is no penalty to choosing additional bins, as long as one does not bin so finely that detector resolution becomes an important issue.  For the analysis with $M_{Z'} = 1.5$ TeV, we choose the bins $0 < |Y| < 0.4$, $0.4 < |Y| < 0.8$, $0.8 < |Y| < 1.2$, and $1.2 < |Y| < Y_{max}$, where $Y_{max} = 1/2 \log(s/M_{Z'}^2)$ is the largest rapidity available.  We later study the effect of the coarser two-bin analysis, and find that only slight improvement is obtained by this finer binning.

\item \underline{Invariant Mass}: Previous studies have examined the above observables and simulated extraction of coupling information at the LHC~\cite{ Petriello:2008zr,delAguila:1993ym,Dittmar:2003ir,Cvetic:1995zs}, or bounded couplings at the Tevatron~\cite{Carena:2004xs}, by integrating the invariant mass around the $Z'$ resonance peak.  Here, the $b$ and $c$ terms in Eq.~(\ref{crs-sect}) can be safely ignored, but any information they might provide is discarded also.  Only squares of couplings are accessible on the resonance peak.  We aim to gain sign information by probing regions where the $Z'$ interference terms with the other gauge bosons have an effect, i.e., we want to study the $b$ terms.  We will do this by including invariant mass bins between the $Z$ pole and the $Z'$ pole.  In the $1.5$ TeV case, we take the bins $800$ GeV $< M < 1000$ GeV, $1000$ GeV $< M < 1200$ GeV, $1200$ GeV $< M < 1400$ GeV, and an on-peak bin.  The on-peak bin should be chosen to contain most of the resonance; three or more widths around the peak should be sufficient.  However, since we assume no knowledge of the width, we must fix a number beforehand.  We have chosen the on-peak bin $1400$ GeV $< M < 1600$ GeV, but the exact size does not affect the quality of the analysis, so long as it is consistent.  For simplicity we refer to the bin as {\it on-peak} henceforth.  
The appropriate size to use for the on-peak bin can be determined with only a rough idea of the $Z'$ resonance shape.
The other bins should be chosen far enough away from the pole that the $a$ terms do not dominate.  For our considered models and mass, observables in each invariant mass bin receive contributions roughly equal in magnitude from the $a, b$, and $c$ terms, with $c$ being somewhat stronger in low mass bins and $a$ being stronger in high mass bins. 

\end{itemize}

\begin{table}[ht]
 \begin{center}
  \begin{tabular}{| c || c | c | c | c |}
\hline
&  Bin 1 & Bin 2 & Bin 3 & Bin 4 \\ \hline\hline
$|Y|$ & $[0,0.4]$ & $[0.4,0.8]$ & $[0.8,1.2]$ & $[1.2,Y_{max}]$ \\ \hline
$M$  (TeV)& $[0.8,1.0] $ & $[1.0,1.2]$ & $[1.2,1.4]$ & \it{on-peak} \\ \hline
$\text{cos}\,\theta^*$ & $[-1,0]$ & $[0,1]$ & $-$& $-$ \\ \hline
  \end{tabular}
 \end{center}
\caption{Summary of bin endpoints used in our analysis for the studied variables: the rapidity $Y$, the invariant mass $M$, and the leptonic scattering angle $\text{cos}\theta^*$.  Each region is displayed 
in the format $[lower,upper]$.  Further details are given in the text.}
\label{bins}
\end{table}

\begin{table}[ht]
 \begin{center}
  \begin{tabular}{| c | c | c || c | c | c | c | c |}
\hline
 \multicolumn{3}{|c||}{} & $SM$ & $\chi$ & $LR$ & $\psi$ & $\chi^*$ \\ \hline \hline
\multicolumn{3}{|c||}{$\sigma_{on-peak}$} & 0.783  & 52.6  & 59.7  & 26.6  & 94.8   \\ \hline
$F$ & $0 < Y < 0.4$ & $800 < M < 1000$ & 1.57  & 1.68  & 1.45  & 1.59  & 1.85   \\ \hline
$B$ & $0.4 < Y < 0.8$ & $1000 < M < 1200$ & 0.382  & 0.524  & 0.464  & 0.378  & 0.772  \\ \hline
$F$ & $0.8 < Y < 1.2$ & $1200 < M < 1400$ & 0.284  & 0.395  & 0.420  & 0.383  & 0.696  \\ \hline
$F$ & $Y > 1.2$ & {\it on-peak} & 0.075  & 1.90  & 3.92  & 1.59  & 3.39  \\ \hline
$B$ & $Y > 1.2$ & {\it on-peak} & 0.025  & 2.95  & 2.52  & 1.52  & 5.33 \\ \hline
$F$ & $0 < Y < 0.4$ & {\it on-peak} & 0.147  & 9.60  & 11.34  & 4.65  & 17.29  \\ \hline
$B$ & $0 < Y < 0.4$ & {\it on-peak} & 0.114  & 10.65  & 10.24  & 4.61  & 19.22  \\ \hline
   
  \end{tabular}

 \end{center}
\caption{Selected expected cross sections for the considered models and the Standard Model, in fb.  Parameters and selection cuts are discussed in the text.}
\label{meas}
\end{table}

In total, we simulate $2 \times 4 \times 4 = 32$ measurements for the main part of our LHC analysis.  A summary of the bins used for $Y$, $M$, and $\text{cos}\,\theta^*$ are shown 
in Table.~\ref{bins}.  For illustrative purposes a selection of numerical results for both the SM and the test models 
using the factorization and renormalization scale choices $\mu_F=\mu_R=M_{Z'}$ can be found in Table~\ref{meas}.

\subsection{Simulation details}

To simulate observation of a $Z'$ at the LHC we perform a fully differential next-to-leading-order (NLO) QCD calculation of all observables described above.  We use CTEQ 6.6 PDFs~\cite{Nadolsky:2008zw}.  The QCD corrections to the cross section are needed to obtain the proper normalization of the observables~\cite{Petriello:2008zr}.  We impose the following basic acceptance cuts on the final state lepton transverse momenta and pseudorapidities: $p_T^l > 20$ GeV and 
$|\eta^l| < 2.5$.  Together, these constraints result in acceptances of roughly 90\%~\cite{Petriello:2008zr}.  Previous studies have found that detector resolution effects and other measurement errors are unlikely to have a significant effect on the 
$e^+e^-$ final state~\cite{Dittmar:2003ir}, and are neglected.  A CMS simulation of $Z'$ production found reconstruction efficiencies near 90\% in the electron channel and no significant detector systematic errors~\cite{Clerbaux}.  In addition, electron energies can be measured to better than 1\% accuracy, and invariant masses can therefore be reconstructed very well.  Reconstruction 
efficiencies above 90\% and invariant mass measurements to sub-1\% precision have also been found for high invariant-mass muon production at the LHC~\cite{muons}.  We 
include here only the statistics for the $e^+ e^-$ production, without reconstruction efficiencies imposed.  In light of the possibility for a factor of two increase in statistics from precision measurements in the muon-pair channel, we believe our study conservatively estimates the LHC potential.  Equivalently, one can rescale all required luminosities by a factor of 2 if including this channel, or use it to separately extract electron couplings and muon couplings.

To construct these measurements, we evaluate Eq.~(\ref{crs-sect}) in the following manner.  The considered models have narrow widths.  For computational efficiency, we calculate the $a$ coefficients once using the $\chi$ width of 17.9 GeV.  We choose this width because it is centrally located among those considered.  On the resonance peak, observables scale as $1/\Gamma$ in the narrow-width approximation.  Since we are not integrating over all invariant masses, we must improve this approximation for a finite bin size denoted $B$.  
When we scan over the width, we scale the $a$ coefficients integrated over the on-peak bins by replacing
\begin{equation}
\int_{(M-B/2)^2}^{(M+B/2)^2} dM^2 a(\Gamma) \ra \frac{\Gamma_{\chi}}{\Gamma}(1 + 2 \frac{\Gamma_{\chi}-\Gamma}{\pi B}) \int_{(M-B/2)^2}^{(M+B/2)^2} dM^2 a(\Gamma_{\chi}) 
\label{nwa_imp}
\end{equation}
to integrate $a(\Gamma)$ for other widths.  Here we have included an $O(\Gamma_{Z'}/B)$ correction, and dropped terms of $O(B^2/M_{Z'}^2)$.  For $M = 1.5$ TeV, we have taken $B = 200$ GeV.  The remaining integral of $a(\Gamma_{\chi})$ is computed using the actual propagator, {\it not} with the narrow-width approximation.  This is accurate to within tenths of a percent over the range of widths considered, and allows the $a$ coefficients to be computed once rather than for all widths scanned over.  If the finite bin-size corrections are completely dropped, errors of the size $5-10\%$ are introduced.   We can then calculate the integrated coefficients $a,b,c$ once the mass is known.   We take $c$ to be the SM Drell-Yan cross section.    

\subsection{Early LHC running \label{early_meas}}

We will also consider subsets of the above measurements for use in discriminating models with early LHC data assuming$\sqrt{s}=10$ TeV.  Our restriction in the number of bins is motivated by the desire to avoid low-count statistical issues and the need to account for detector miscalibration, which would over-complicate our analysis.  At low luminosities the reduction in bins does not sharply reduce our sensitivity to differences in the underlying model parameters. We merge the bins as follows for 10 TeV running.
\begin{itemize}
 \item \underline{30, 100 fb$^{-1}$}:
The two highest off-peak invariant mass bins are merged (now $1000 < M < 1400$), as are the two highest rapidity bins (now $Y > 0.8$), for $2 \times 3 \times 3 = 18$ total bins.  While the LHC will hopefully increase in energy prior to reaching this integrated luminosity, we consider these measurements at 10 TeV as a bridge between the early running and the late/SLHC running.
\item \underline{10 fb$^{-1}$}:
All off-peak invariant mass bins are merged ($800 < M < 1400$), as are the two highest rapidity bins, for $2 \times 3 \times 2 = 12$ total bins.

\item For the pairwise model comparison at a few femtobarns of integrated luminosity to be described in Section~\ref{sec:LHCanalysis}, we consider only measurements on the resonance peak.  For simplicity we use four bins: 2 rapidity bins ($|Y| < 0.6, |Y| > 0.6$), and the forwards/backwards measurement.  

\end{itemize}

Below we orient the reader by providing 10 TeV on-peak cross sections for our test models.  The parton luminosity for $M_{Z'} = 1.5$ TeV is consistently reduced by approximately a factor of 2.5 from $\sqrt{s} = 14$ TeV.
\begin{table}[hbt]
 \begin{center}
  \begin{tabular}{| c | c |}
\hline
Model & $\sigma$ (fb) \\ \hline
$\chi$ & 20.5 \\ \hline
LR & 23.8 \\ \hline
$\psi$ & 10.7 \\ \hline
$\chi^*$ & 36.9 \\ \hline
   
  \end{tabular}

 \end{center}
\caption{On-peak cross sections for test models at $\sqrt{s} = 10$ TeV.}

\end{table}

\section{Low-energy Measurements}
\label{sec:lowmeas}

The upgrade of the Jefferson Lab polarized $e^-$ beam will advance the precision frontier for low-energy parity violation.  The Q-weak experiment~\cite{VanOers:2007zz} will probe the effective weak charge of the proton in $ep$ scattering, and the proposed M\o{}ller scattering experiment~\cite{mollerup} can determine $\sin^2\theta_W^{eff}$ through a precision measurement of the 
asymmetry in $e^-e^-$ scattering.  Deviations from predicted SM values due to a $Z'$ can provide further coupling information once the $Z'$ mass is known from the LHC.  We describe here which 
$Z'$ parameters to which each experiment is sensitive, and review the expected errors on the measured quantities arising from both experimental issues and SM theoretical uncertainties.

\subsection{Q-weak}
The Q-weak experiment will probe the effective weak charge of the proton, $Q_W^p \simeq 1 - 4 \sin^2 \theta_W$, by observing the parity-violating asymmetry in $ep$ scattering.  Due to the smallness of $1 - 4 \sin^2 \theta_W$, higher-order corrections are important.  The predicted low-energy value in the SM (from running from $Z$-pole measurements) is~\cite{Erler:2004in}
\be
Q_W^p = 0.0713 \pm 0.0008 .
\ee
The anticipated experimental error after Q-weak running is 4.3\%~\cite{VanOers:2007zz}.  The experimentally observed value will differ due to the presence of a $Z'$; the shift is~\cite{RamseyMusolf:1999qk}
\be
\Delta Q_W^p = \frac{2 \sqrt{2}}{M_{Z'}^2 G_F} (2 e_A u_V + e_A d_V) ,
\ee
where $A$ and $V$ denote the axial-vector and vector charges, respectively.  We rewrite this in terms of the parameters defined in Section~\ref{params} to derive
\be
\Delta Q_W^p = \frac{\sqrt{2}}{2 M_{Z'}^2 G_F} (3 q_L + 2 u_R + d_R)(e_R - e_L) .
\label{qweakshift}
\ee
Like LHC observables in the Drell-Yan channel, the Q-weak measurement is sensitive only to the coupling combination $q\times e$.

\subsection{M\o{}ller scattering}

The goal of the proposed J-Lab M\o{}ller experiment is to measure the effective weak mixing angle, $\sin^2 \theta_W^{eff}$, to an absolute precision of 0.00025, or in terms of the weak charge of the electron, to within 2.3\%.  The running of $\sin^2 \theta_W$ in the SM gives a prediction for this quantity at low energies~\cite{Erler:2004in}.  The result depends rather strongly on the scale at which $\gamma Z$ box diagrams are evaluated: $Q_W^e(0) = -0.0472$, while $Q_W^e(M_Z) = -0.0462$~\cite{Erler_priv}.  Higher-order calculations are needed to determine the appropriate scale 
at which to evaluate these contributions.  Following the suggestion of~\cite{Czarnecki:1995fw}, we split the difference and take it as a contribution to the theoretical uncertainty:
\be
Q_W^e & = & 0.0467 \pm 0.0006 (Z \rm{- pole}) \pm 0.0005 (\gamma Z) \pm 0.0011 (\rm{experiment}) \\
      & = & 0.0467 \pm 0.0013,
\ee
where we have added the various sources of error in quadrature.  Shifts from the expected asymmetry, similarly to Q-weak, go as
\begin{align}
\Delta Q_W^e & = \frac{2 \sqrt{2}}{M_{Z'}^2 G_F} e_A e_V \notag \\
	& = \frac{\sqrt{2}}{2 M_{Z'}^2 G_F} (e_R^2 - e_L^2) .
\end{align}
An interesting feature of this quantity is that it depends only on electron couplings, and not those of the quarks.  A measurement of this quantity would break 
the scaling degeneracy that plagues the other measurements discussed previously.

\section{LHC and Q-weak Analysis}
\label{sec:LHCanalysis}

To extract coupling information from the LHC from the measurements described in Section~\ref{sec:LHCmeas}, we must evaluate the coefficients $a, b$, and $c$ in Eq.~(\ref{crs-sect}).  This requires knowing the $Z'$ mass from experiment.  For illustrative purposes we explore the possibility $M_{Z'} = 1.5$ TeV, and take the mass to have negiligble error~\cite{atlas_sim}.  Since we are not making any assumptions on how well the width is measured directly, we must treat this as a free parameter in our analysis, and fit it accordingly.  This leaves five unknown parameters on which Eq.~(\ref{crs-sect}) depends: $\Gamma_{Z'}, q_Le_L, q_Le_R, u_Re_L,$ and $d_Re_L$.  Two other combinations, $u_Re_R$ and $d_Re_R$, appear in Eq.~(\ref{crs-sect}); however, these are not independent from those listed due to the following relations:
\be
\frac{q_Le_R}{q_Le_L} = \frac{u_Re_R}{u_Re_L} = \frac{d_Re_R}{d_Re_L} = \frac{e_R}{e_L} \equiv \tan \theta_l .
\ee
Here we have defined the parameter $\theta_l$, which is the angle in the plane of the leptonic couplings.  There is no way to separate $q$ couplings from $e$ couplings at this stage, as the $Z'$ is produced via quarks and decays into leptons in this channel.  This limitation is addressed in the next section.  We include a discussion of the Q-weak measurement in this section as it probes 
the same $q \times e$ combination as the LHC.

We determine the five unknown parameters by performing a scan over this parameter space.  For each point in the five-dimensional parameter space, we can reconstruct the predicted LHC measurement bins discussed in Sec.~\ref{sec:LHCmeas} using Eq.~(\ref{crs-sect}), and compare to the actual observed values.  In the absence of data, we construct measurements for the $\chi, \psi$, and LR models described in Section~\ref{models} to test our procedure.  As we present analyses for several coupling combinations with varying assumptions regarding integrated luminosity and energy, we organize our 
results into subsections as follows.
\begin{itemize}

\item We first present a discussion of the relevant uncertainties affecting both the LHC and Q-weak measurements.

\item We discuss an initial three-parameter $\chi^2$ comparison between data and model hypotheses that can be performed with only a few ${\rm fb}^{-1}$ of data at the LHC.

\item We present results for integrated luminosities between $10-100\;{\rm fb}^{-1}$ for the $c_q$ and $e_q$ coupling combinations defined in~\cite{Petriello:2008zr}, which are most relevant when only on-peak data gives significant information.

\item We then discuss extraction of the $q \times e$ coupling combinations assuming large LHC integrated luminosities of $300-1000 \; {\rm fb}^{-1}$, and demonstrate how the $Z'$ width can 
be determined by comparing on-peak and off-peak bins.  We add the expected Q-weak measurement to the analysis at this stage.

\item Finally, we summarize our analysis and present results showing how well all $Z'$ parameters relevant at the LHC can be determined as a function of integrated luminosity.

\end{itemize}

\subsection{LHC errors}
Our ultimate goal is to extract the underlying $Z'$ couplings.  We will need to perform a $\chi^2$ test between hypothesis points in the coupling parameter space and the ``observed'' values for each of our example models.  For early LHC running, we will want to know whether one can start eliminating canonical models, and we do a pairwise comparison between our test models to see how well this can be done.  For later LHC running, we will demonstrate the efficacy of our coupling extraction procedure.  To do so we must study the uncertainties which affect our analysis procedure.  We classify errors into four categories.
\begin{itemize}

\item \underline{PDF}.  We determine the PDF errors using the CTEQ 6.6 PDF error sets~\cite{Nadolsky:2008zw} according to the prescription of~\cite{Pumplin:2002vw}, and account for correlations among our measurement bins.  To be conservative, we take the larger magnitude shifts from each of the 22 eigenvector directions in PDF parameter space, and when forming the $\chi^2$, interpret these errors as 1$\sigma$.  If the $Z$ cross section is used as a standard candle, one should instead find the PDF error in the ratios of our bins to the expected number of $Z$ events.  If there were significant correlations between our bins and the $Z$ cross section, this would reduce PDF errors.  Testing indicates this is not the case, and for simplicity we do not normalize to $Z$ production to determine PDF errors. 

\item \underline{Statistical}.  For each considered value of integrated luminosity we determine the counting error in each bin.  The bin choices for each luminosity are described in Section~\ref{sec:bins}.

\item \underline{Theory}.  We have calculated all cross sections at NLO in the strong coupling.  Residual scale errors are negligible compared to our other errors~\cite{Petriello:2008zr}, 
and it is known that the NNLO corrections have a very small impact on the central value of the NLO result~\cite{SMDY,Coriano:2008wf}. QED corrections in the final state may be important in predicting the invariant mass distribution~\cite{Baur:1997wa, Baur:2001ze}, though with our coarse binning this issue is less important.  Further electroweak corrections come in two forms: vertex corrections, which can be absorbed into the effective couplings extracted, and box corrections (where two sides of the box are $W, Z$, or $Z'$), which cannot.  The latter have large logarithmic corrections in the energy, $\log(\hat{s}/M_{W,Z}^2)$ \cite{Ciafaloni:1998xg}, and should be determined when examining off-peak bins.  We assume that a full analysis will have computed these corrections; the actual shifts in the coefficients $a, b$, and $c$ should not be large enough to affect the efficiency of our method.  However, we do include, conservatively, a 5\% error in the overall normalization of the off-peak bins to account for remaining errors {\it after} EW corrections have been included.  We have found that including this error does not substantially affect the quality of the coupling extraction.

\item \underline{Detector}.  Our binning is coarse enough that we take detector issues such as energy mis-measurement, cracks, etc., to be sub-dominant.  We neglect them in our study.

\end{itemize}

\subsection{Q-weak errors}
The Q-weak experiment will also probe combinations of quark $\times$ lepton couplings; from Eq.~(\ref{qweakshift}), it will measure the combination

\be
(3 q_Le_L + 2 u_Re_L + d_Re_L)(\frac{q_Le_R}{q_Le_L} - 1) = \sqrt{2} M_{Z'}^2 G_F \Delta Q_W^p.
\ee
The $Z'$ mass will be known from the LHC.  The anticipated uncertainty after combining both experimental and theoretical errors is large compared to the expected deviations of our test models:

\be
\delta \frac{\Delta Q_W^p}{Q_W^p} = 4.4\%, \\
\delta \Delta Q_W^p = 0.0032
\ee
For reference, we list the expected shifts due to our test models at 1.5 TeV in Table~\ref{qweaknums}.  A $Z'$ would have to be fairly strongly coupled to be observable by Q-weak.  The Q-weak experiment may add somewhat to the discriminating power of the LHC, however.  We assume that Q-weak finds a value consistent with our test model, and for each test point, we generate a measurement and add it to the $\chi^2$ appropriately.  For simplicitly, the $Z'$ parameters are not run from the $Z'$ scale down, but compared directly.  If the running is $O(10\%)$, as expected, this can be safely neglected compared to the experimental error.

\begin{table}[htbp]
 \begin{center}
  \begin{tabular}{| c | c |}
\hline
& $\Delta Q_W^p \times 10^3$ \\ \hline
$\chi$ & 1.17 \\ \hline
$LR$ & -0.45 \\ \hline 
$\psi$ & 0 \\ \hline
$\chi^*$ & 2.24 \\ \hline
  \end{tabular}

 \end{center}
\caption{Expected shifts in the weak charge due to a 1.5 TeV $Z'$, in unit of $10^{-3}$.  This should be compared to the expected measurement precision of 
$\delta \Delta Q_W^p = 0.0032$. \label{qweaknums}}
\end{table}

\subsection{Early LHC results: $\chi^2$ comparisons}
\label{sec:chi2}

We first assume a center-of-mass energy of only $\sqrt{s} = 10$ TeV, consistent with early-running scenarios.  At low luminosities, the statistical errors are too large for a meaningful coupling extraction.  Therefore we ask the question, is it possible to begin to discriminate models?  How can one rule out a ``model''?  High-scale models typically determine the $Z'$ fermion charges, but the overall coupling running or the leptonic branching fraction may depend on other parameters of the theory not considered here, or even predicted by the model.  Therefore, in this first pass to distinguish models, we wish to consider observables independent of these parameters.  We first note the following.  For very early running, the error in off-peak observables is too large for them to be useful.  Secondly, the overall on-peak cross section, $\sigma_{peak}$, is proportional to the overall coupling squared and the branching fraction: $\sigma_{peak} \propto g_{Z'}^2 Br(Z' \rightarrow e^+ e^-)$.  Therefore, if one takes on-peak observables normalized by the overall on-peak cross section, $\hat{\sigma}_i = \sigma_i / \sigma_{peak}$, they are independent of the overall coupling and leptonic branching fraction.  Considering these observables only will allow us to differeniate models defined only up to these ambiguous parameters.  In forming our $\chi^2$ comparison using the four bins described in Sec.~\ref{early_meas}, we normalize the bins to the on-peak cross section, and we restrict ourselves to on-peak bins.  As these observables are no longer independent ($\sum_i \sigma_i = \sigma_{peak}$), one bin is dropped, for three observables, and three degrees of freedom.  We present the $\chi^2$ comparison of experimentally ``found'' modes versus other model hypotheses in Table~\ref{chicomp}.
For reference, 68\% confidence corresponds to $\chi^2 = 3.5$, 90\% to 6.3, and 95\% to 7.8.

\begin{table}[ht]
 \begin{center}
  \begin{tabular}{| c | c c c | c c c | c c c |}
   \hline
 & \multicolumn{3}{| c |}{3 fb$^{-1}$} & \multicolumn{3}{| c |}{10 fb$^{-1}$} & \multicolumn{3}{| c |}{30 fb$^{-1}$} \\ \hline
 & $\chi$ & $\psi$ & LR & $\chi$ & $\psi$ & LR & $\chi$ & $\psi$ & LR \\ \hline
$\chi$ & - & 1.7 & 6.7 & - & 5.1 & 23.1 & - & 15.7 & 54.7 \\
$\psi$ & 3.8 & - & 0.58 & 11.3 & - & 3.3 & 30.6 & - & 5.2 \\
LR & 6.8 & 0.3 & - & 22.9 & 1.4 & - & 55.1 & 2.6 & - \\ \hline

  \end{tabular}

 \end{center}
\caption{$\chi^2$ comparison (with three degrees of freedom) between models for early LHC running.  Hypothesis models are in columns, and are tested against the experimentally found row models.  Errors are determined from the hypothesis model.}
\label{chicomp}

\end{table}

In performing this test, we have assumed statistics consistent with our prior assumptions about the overall coupling and width for the experimentally ``found'' model; we remind the reader that our procedure is independent of these assumptions.  We note that the $\psi$ model is difficult to differentiate under these assumptions due to its lower production rate.  We note that ``nice'' models with clean signatures such as the $\chi$ model and left-right model can be distinguished from each other at over 90\% confidence with as little as 3 fb$^{-1}$.

\subsection{Early LHC results: initial parameter extractions}
\label{sec:initlhc}

\begin{figure}[htbp]
 \centering
\includegraphics[scale=0.60,angle=90]{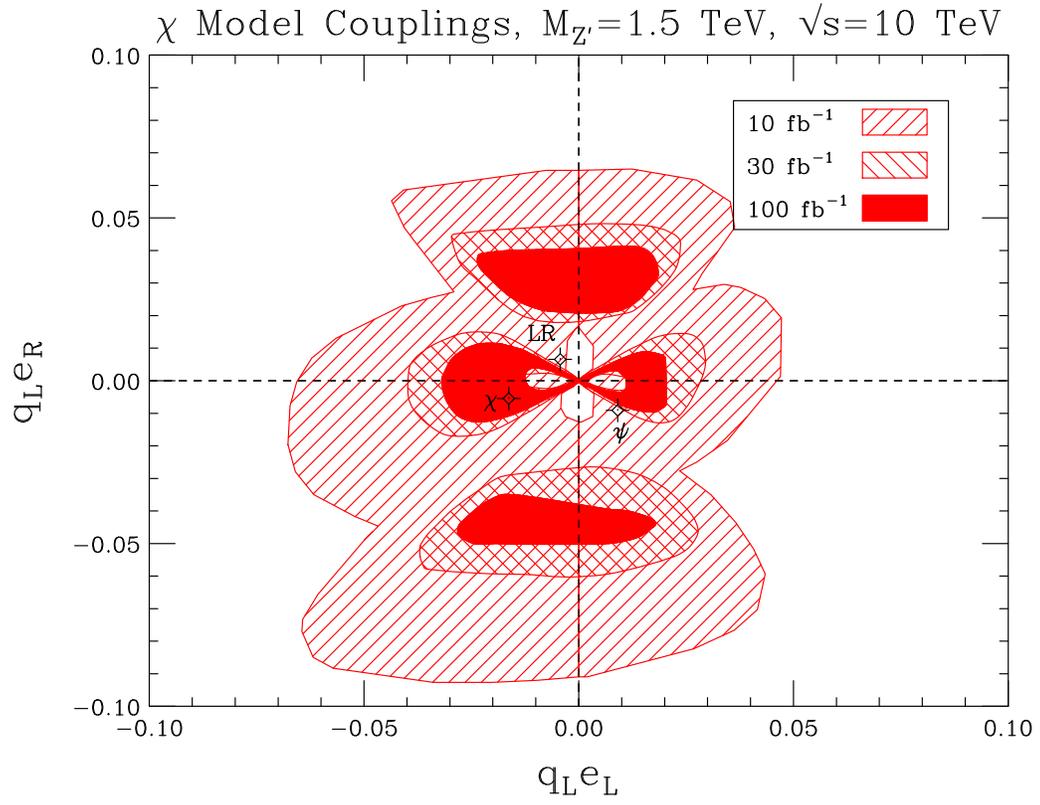}
\caption{95\% confidence-level region for the $q_L\times e_R$ and $q_L \times e_L$ couplings extracted from measurements corresponding to the $\chi$ model, at 10, 30, and 100 fb$^{-1}$.  Other model values for these couplings are shown for orientation.  \label{early_coup}}
\end{figure}

For slightly higher luminosities, we can use the off-peak bins to begin to reduce the allowed region in coupling space.  As an example, we plot the 95\% confidence region for two of the coupling combinations ($q_L\times e_R$ and $q_L \times e_L$) for measurements corresponding to the $\chi$ model in Fig.~\ref{early_coup}.  If all couplings and the width are fit simultaneously, this corresponds to a 68\% confidence region.  We make the following observations.
\begin{itemize}
 \item At 10 fb$^{-1}$, one only gets a vague notion of the size of the couplings.  Since we do not assume the width is known from the Breit-Wigner scan at this stage, this is not due to knowledge of the on-peak total cross section.  It is instead coming from the limited size of the deviations of the off-peak bins from SM predictions.
\item At 30 fb$^{-1}$, we see multiple ``islands'' emerge; these are due to the nonlinearity of the cross section in the $q \times e$ parameters.  The allowed coupling space has shrunk considerably, and we see the beginnings of a measurement for the correct island.  Note, however, that all sign degeneracies remain in the couplings; it is still the on-peak measurements leading to reduction of the 
allowed region.  To determine these signs, one must wait for more integrated luminosity.
\end{itemize}

\begin{figure}[htbp]
 \centering
\includegraphics[scale=0.44,angle=90]{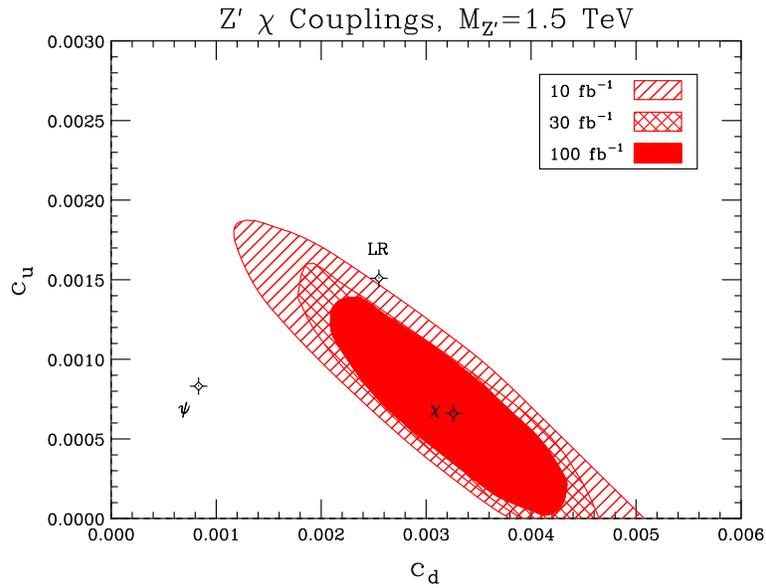}
\includegraphics[scale=0.44,angle=90]{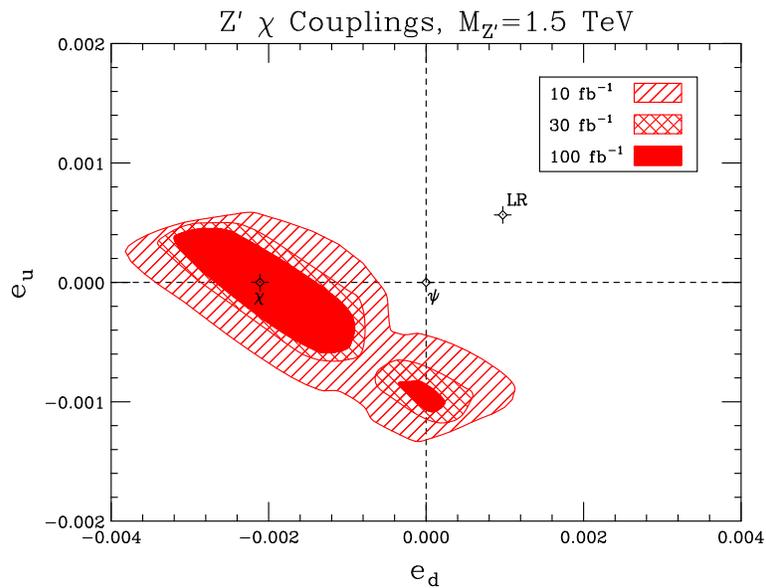}
\caption{68\% confidence regions for the coupling combinations $c_q$ and $e_q$ extracted from measurements corresponding to the $\chi$ model, at 10, 30, and 100 fb$^{-1}$.  Other model values are shown for orientation.  We note that $c_q$ and $e_q$ also have correlations with each other that are not displayed.  \label{c_e_2d}}
\end{figure}

At this stage, the couplings themselves are not measured particularly well, though we are seeing quite a reduction in allowed parameter space.  In Ref.~\cite{Petriello:2008zr}, the anticipated errors for the combinations of the couplings measurable on peak were derived.  Given the success of that extraction, we wish to revisit these results using the chosen binning and lower luminosity, again choosing the $\chi$ model as an example.  We plot in Fig.~\ref{c_e_2d} the 68\% confidence regions for the four parameters $c_q$ and $e_q$ ($q=u,d$), defined in Ref.~\cite{Petriello:2008zr} as

\begin{eqnarray}
 c_q &= & (q_R^2+q_L^2)(e_R^2+e_L^2)\frac{M_{Z'}}{24 \pi \Gamma_{Z'}}, \nonumber \\
 e_q &= & (q_R^2-q_L^2)(e_R^2-e_L^2)\frac{M_{Z'}}{24 \pi \Gamma_{Z'}} .
\label{cedef}
\end{eqnarray}
These have the advantage that they are directly related to the cross section near the resonance peak,
\be
\frac{d^2\sigma}{dYd\cos\theta^*dM^2} \simeq \sum_{q=u,d} & [a_1^{'q}(M_{Z'}, \Gamma_{Z'})c_q + a_2^{'q}(M_{Z'}, \Gamma_{Z'})e_q] \label{peak_crs_sect}
\ee
where we have rescaled the coefficients $a_i^{'q} = \frac{24 \pi \Gamma_{Z'}}{M_{Z'}} a_i^q $ of Eq.~(\ref{crs-sect}) to move most of the width dependence into the coupling combinations $c_q$ and $e_q$.  For this analysis, since we use a fixed bin size, we also fold a factor of $(1 + 2 \frac{\Gamma_{\chi}-\Gamma}{\pi B})$ from Eq.~(\ref{nwa_imp}) into the definitions of $c_q$ and $e_q$ to keep the results practically width-independent.  This is relatively unimportant if the width is small relative to the on-peak bin size, or known a priori.

The on-peak bins are doing almost all the work discriminating $c_q$ and $e_q$, as these quantities are directly related to the on-peak bins.  The off-peak bins do not significantly help in reducing the allowed coupling space at these luminosities.  As such, we see nearly elliptical errors as one would expect from a linear propagation of errors from Eq.~(\ref{peak_crs_sect}).  However, the errors differ slightly from our previous analysis~\cite{Petriello:2008zr}, and we wish to mention the following differences.
\begin{itemize} 

\item These are 68\% confidence regions for 2 parameters, and not $1 \sigma$ in each parameter.

\item $\sqrt{s} = 10$ TeV, not 14 TeV, which affects both statistical and PDF errors.

\item The physical constraint $u_L = d_L$ is explicit in the parameter space scan, as are $c_q > 0$, $|e_q| < c_q$, etc.

\item CTEQ 6.6 PDFs are used instead of CTEQ 6.5, which allows for both $s$ quark differences and improves high $x$ errors~\cite{Nadolsky:2008zw}.

\item Less conservative assumptions are made on the correlations of differing PDF eigenvector directions.

\end{itemize}

We note that under these assumptions $u$ and $d$ coupling measurements are very anti-corellated on peak; the sum is known much better.  This is to be expected; their separation depends on subtracting low and high rapidity $Z'$ bins, thereby increasing errors, while the sum depends on combining bins.

\subsection{Late LHC and Q-weak results}

\begin{figure}[htbp]
 \centering
\includegraphics[scale=0.6,angle=90]{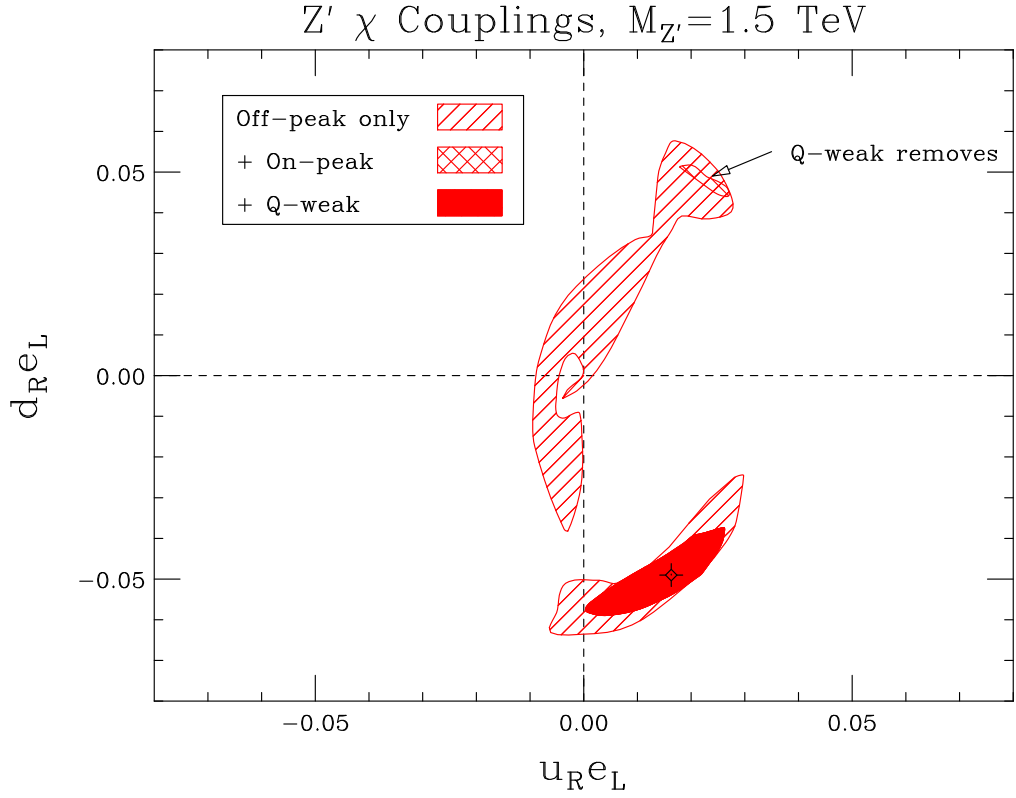}
\caption{Reduction in allowed coupling space with successive application of off-peak measurements only, on-peak measurements, and the Q-weak expected result for the $\chi$ model, at 95\% CL.  We note the removal of the small island in the upper-right quadrant upon the inclusion of Q-weak.  These results include 1000 ${\rm fb}^{-1}$ of integrated luminosity at the LHC.  \label{sucimp}}
\end{figure}

We finally consider the improvements possible if integrated luminosities of 300 ${\rm fb}^{-1}$ or 1 ${\rm ab}^{-1}$ are available.  We now add information from a Q-weak measurement of $Q_W^p$, assuming it finds a value consistent with that predicted by a test model.  We demonstrate in Fig.~\ref{sucimp} the successive improvement upon adding different data in our extraction for 1 ab$^{-1}$ at the LHC, by first applying the off-peak bins only to determine the couplings for the $\chi$ model, then adding on-peak bins, and finally the Q-weak experiment.  We choose to show $u_R e_L$ versus $d_R e_L$ to illustrate this improvement.  Due to the nonlinear nature of the cross section in the $Z'$ parameters, multiple islands are possible, where one region of parameter space can mimic the signal of another within the allowed errors.  The Q-weak experiment is helpful in breaking degeneracies such as these that may remain, where the false islands have a different enough value of $\Delta Q_W^p$ to rule them out {\it in conjunction with} the LHC measurements.  The difference is not in general substantial enough to rule out the islands if the LHC did not already disfavor them.

\begin{table}[htbp]
 \begin{center}
  \begin{tabular}{| c | c  c  c | c  c  c | c  c  c |}
\hline
 & \multicolumn{3}{|c|}{$\chi$} & \multicolumn{3}{|c|}{$LR$} & \multicolumn{3}{|c|}{$\psi$} \\ 
 & Act. & LL & HL & Act. & LL & HL & Act. & LL & HL \\ \hline
$\Gamma_{Z'}$ (GeV) & 17.9 & $\pm^{1.9}_{1.8}$ & $\pm^{1.3}_{1.1}$ & 33.8 & $\pm^{1.6}_{1.4}$ & $\pm^{1.1}_{1.1}$ & 8.1 & $\pm^{6.0}_{2.7}$ & $\pm^{2.1}_{1.5}$ \\ \hline
$100 \times q_Le_L$ & -1.63 & $\pm^{0.29}_{0.22}$ & $\pm^{0.16}_{0.14}$ & -0.43 & $\pm^{0.42}_{0.52}$ & $\pm^{0.33}_{0.38}$ & 0.91 & $\pm^{0.75}_{0.83}$ & $\pm^{0.37}_{0.61}$ \\ \hline 
$100 \times q_Le_R$ & -0.55 & $\pm^{0.14}_{0.08}$ & $\pm^{0.09}_{0.06}$ & 0.66 & $\pm^{0.98}_{0.64}$ & $\pm^{0.64}_{0.50}$ & -0.91 & $\pm^{1.35}_{0.50}$ & $\pm^{0.76}_{0.44}$ \\ \hline
$100 \times u_Re_L$ & 1.63 & $\pm^{0.51}_{0.69}$ & $\pm^{0.44}_{0.53}$ & 2.83 & $\pm^{0.24}_{0.48}$ & $\pm^{0.23}_{0.34}$ & -0.91 & $\pm^{0.83}_{0.78}$ & $\pm^{0.67}_{0.37}$ \\ \hline
$100 \times d_Re_L$ & 4.90 & $\pm^{0.62}_{0.60}$ & $\pm^{0.40}_{0.43}$ & -3.70 & $\pm^{0.37}_{0.22}$ & $\pm^{0.20}_{0.16}$ & -0.91 & $\pm^{2.62}_{0.50}$ & $\pm^{1.80}_{0.47}$ \\ \hline
$\theta_l$ & $18.4^{\circ}$ & $\pm^{4.6}_{5.7}$ & $\pm^{3.0}_{3.8}$ & $-56.5^{\circ}$ & $\pm^{0.9}_{3.2}$ & $\pm^{0.7}_{1.7}$ & $-45^{\circ}$ & $\pm^{61}_{42}$ & $\pm^{39}_{32}$ \\ \hline
  \end{tabular}

 \end{center}
\caption{Results of the LHC/Q-weak coupling extraction.  We use the following abbreviations for the integrated luminosity: LL = 300 fb$^{-1}$, HL = 1 ab$^{-1}$. \label{coup_tab}}
\end{table}

We use the Q-weak experiment in conjunction with the LHC at high luminosities to fit the couplings and width for all models.  In Table~\ref{coup_tab} we list the results of the extraction of these parameters for integrated luminosities of 300 fb$^{-1}$ and 1 ab$^{-1}$.  The errors listed are $1 \sigma$ for each individual parameter, while allowing the other values to float.  For brevity we omit the $\chi^*$ model.  Interestingly, the width is measured extremely well by the extraction.  This essentially comes from comparing the size of on-peak bins to what would be predicted by extracting the couplings off-peak only; the size of the on-peak cross section is directly controlled by the width.  We note that the precision of this measurement is not directly controlled by the 
calorimeter resolution, and is therefore controlled by different systematic errors than a fit to the resonance peak.  The errors here are competitive with a previous detector-level analysis~\cite{atlas_sim} of the Breit-Wigner fit with a fixed number of events corresponding to luminosities of 100-300 fb$^{-1}$.  We have tested that adding knowledge of the width from the Breit-Wigner fit following the 
study in Ref.~\cite{atlas_sim} does not significantly improve the quality of our results.  

\begin{figure}[htbp]
 \centering
\includegraphics[scale=0.6,angle=90]{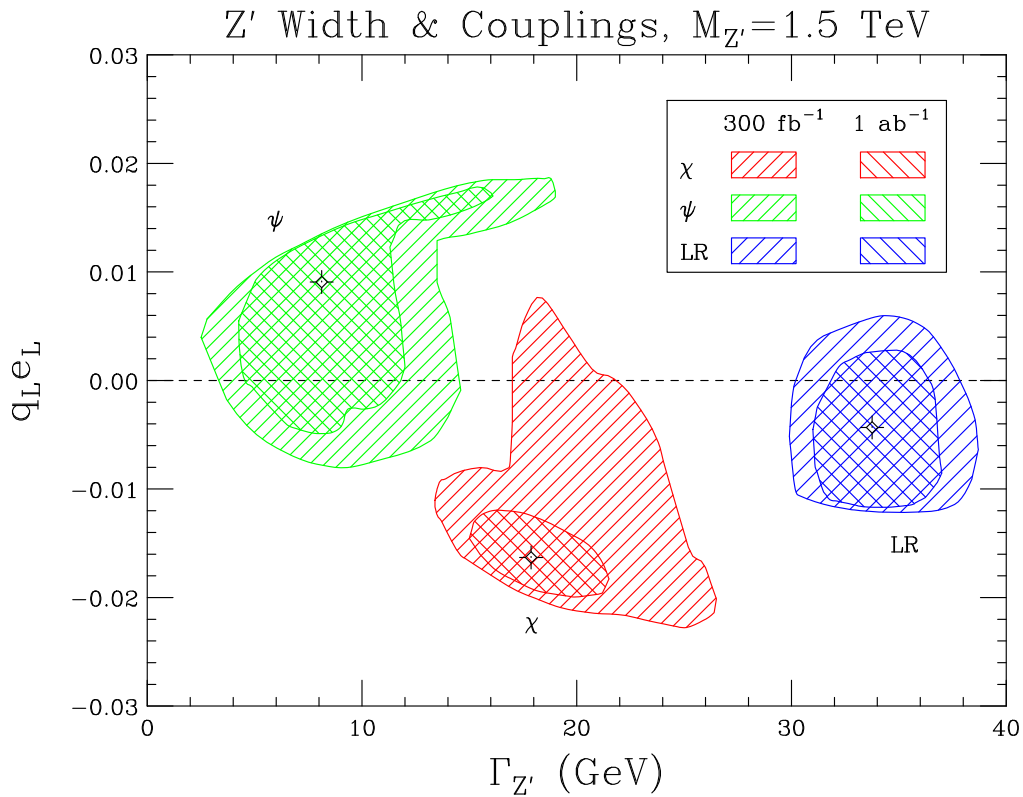}
\caption{Width versus $q_L e_L$ fits for the considered models.  Widths tend to correlate with larger couplings. \label{widthplot}}
\end{figure}

\begin{figure}[htbp]
 \centering
\includegraphics[scale=0.45,angle=90]{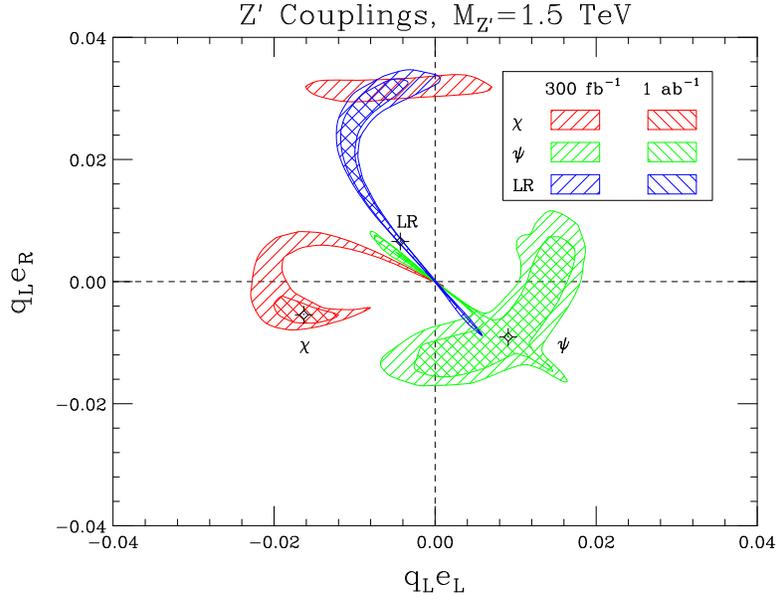}
\includegraphics[scale=0.45,angle=90]{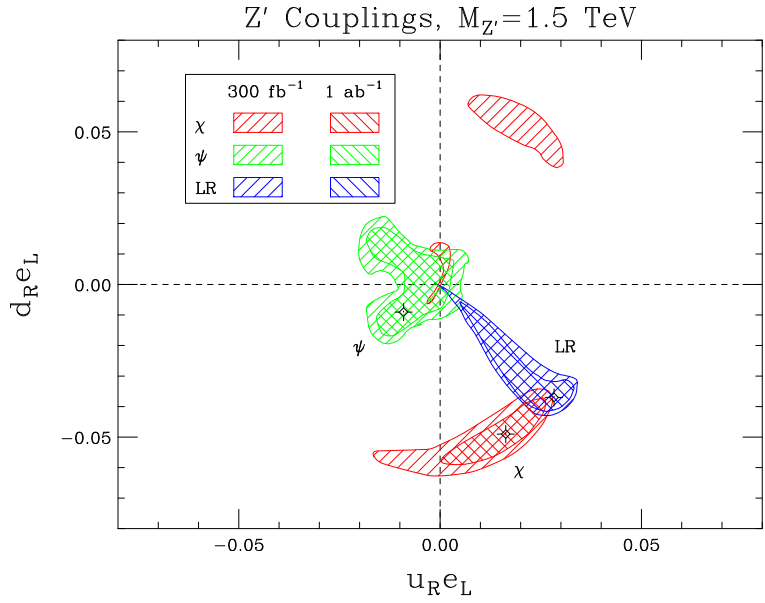}
\caption{Coupling fits for the considered models.  The actual value for each model is shown by the labeled black symbols.\label{coupplots}}
\end{figure}

The couplings, especially the larger ones, are measured to good precision.  The $\psi$ model is an exception; several parameters are measured poorly.  This is due to the lack of asymmetry, where the differences $F - B$ are zero, giving a large relative statistical error.  For non-leptophobic, narrow-width $Z'$ models, we consider this a worst-case scenario.  However, the table above does not tell the whole story.  The 5D 68\% CL region includes strong correlations between the parameters, and portrays a dramatic reduction of parameter space compared to a naive estimate from the table alone.  We plot several 2D projections of the 5D confidence region for our test models in Figs.~\ref{widthplot}~and~\ref{coupplots}  to demonstrate.  These regions can also be interpreted as 95\% confidence regions for the two parameters displayed in each plot separately, allowing the others to float.

We make a few remarks on these results.  First, the $\chi$ model measurement improves substantially with increased (super-LHC) luminosity.  Remaining degeneracies in the parameter space are removed.  However, for the other models, there is only minor improvement.  The allowed volume of coupling space is small, and our test models are easily distinguished.  Also, we point out that it is often possible to mimic the signals of our test models within errors while keeping $q_L$ zero, as can be seen in the upper panel of Fig.~\ref{coupplots}.  If the average asymmetry coming from the quarks is small enough, one can account for it in the leptonic couplings, and get the relative amount of $u$ and $d$ correct by adjusting the values of $u_R$ and $d_R$ without affecting other measurements.  For example, the $\psi$ model is extremely difficult to pin down due to the lack of asymmetries.  One cannot even favor a sign on $d_R e_L$ in this model, as can be seen from the lower panel of Fig.~\ref{coupplots}.

\subsection{Summary: single parameter extractions}
\label{sec:parex}

As our analysis contained many parameters, we review here the combinations that can be probed by the various data sets.  At low luminosity, we have seen that the majority of the analyzing power 
comes from the on-peak bin.  The natural coupling combinations to utilize with this data set are the $c_q$ and $e_q$ combinations defined in Eq.~(\ref{cedef}).  These quantities are formed from 
$(q \times e)^2$.  At higher luminosities, the off-peak bins contain enough events to allow a direct measurement of the linear combinations $q \times e$.  We summarize this information in Table~\ref{sumtab}. Only six of seven possible parameters appear in this table; this reflects the $q \to yq, e\to e/y$ degeneracy discussed earlier.

\begin{table}[htbp]
 \begin{center}
  \begin{tabular}{| c | c |}
\hline
on-peak LHC: & $M_{Z'},c_u,c_d,e_u, e_d$ \\ \hline
on-peak+off-peak LHC: & $M_{Z'},\Gamma_{Z'}, q_Le_L, q_Le_R, u_Re_L,d_Re_L$ \\ \hline
Q-weak: & $(3 q_L + 2 u_R + d_R)(e_R - e_L)/M_{Z'}^2$ \\ \hline
  \end{tabular}
 \end{center}
\caption{Summary of which $Z'$ coupling combinations can be probed with the various data sets considered. \label{sumtab}}
\end{table}

\begin{figure}[htbp]
 \centering
\includegraphics[scale=0.45,angle=90]{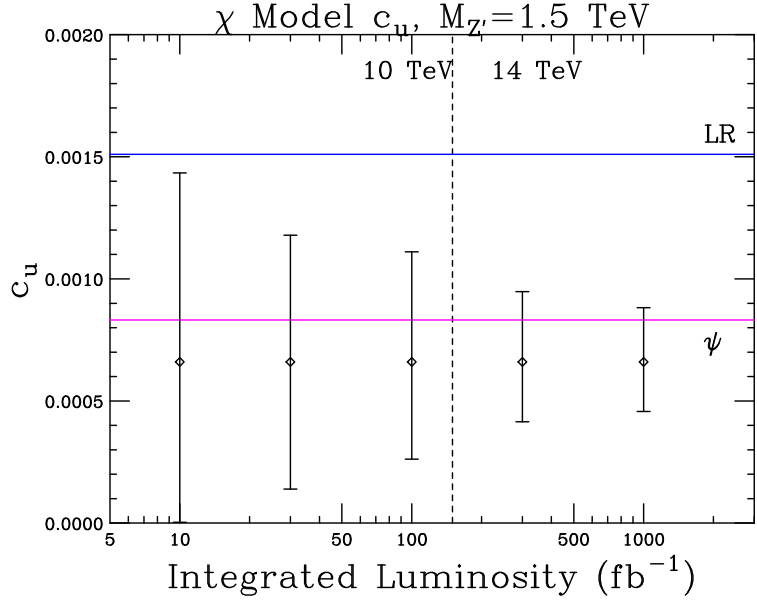}
\includegraphics[scale=0.45,angle=90]{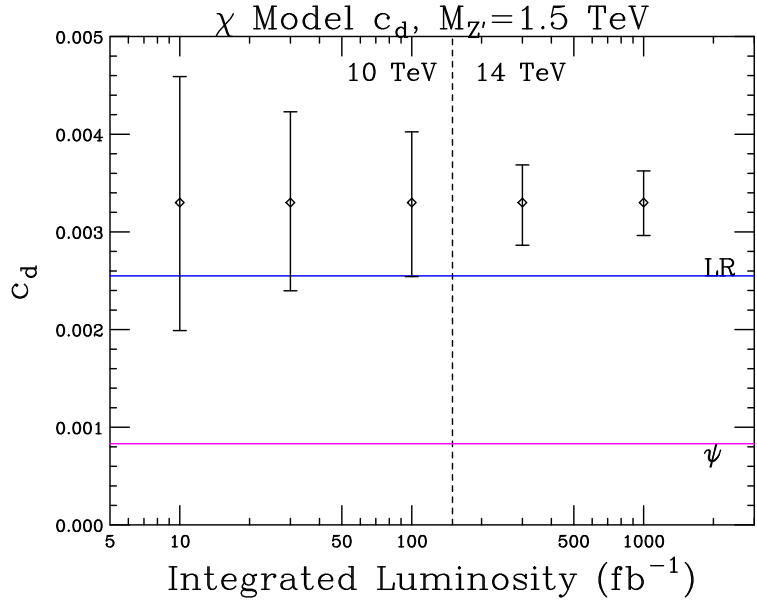} 
\caption{1$\sigma$ errors for the on-peak coupling combinations $c_u$ and $c_d$ as a function of integrated luminosity.  To the left of the dashed line, $\sqrt{s}=10$ TeV, while $\sqrt{s}=14$ TeV 
to its right.
\label{c_1d}}
\end{figure}

\begin{figure}[htbp]
 \centering
\includegraphics[scale=0.45,angle=90]{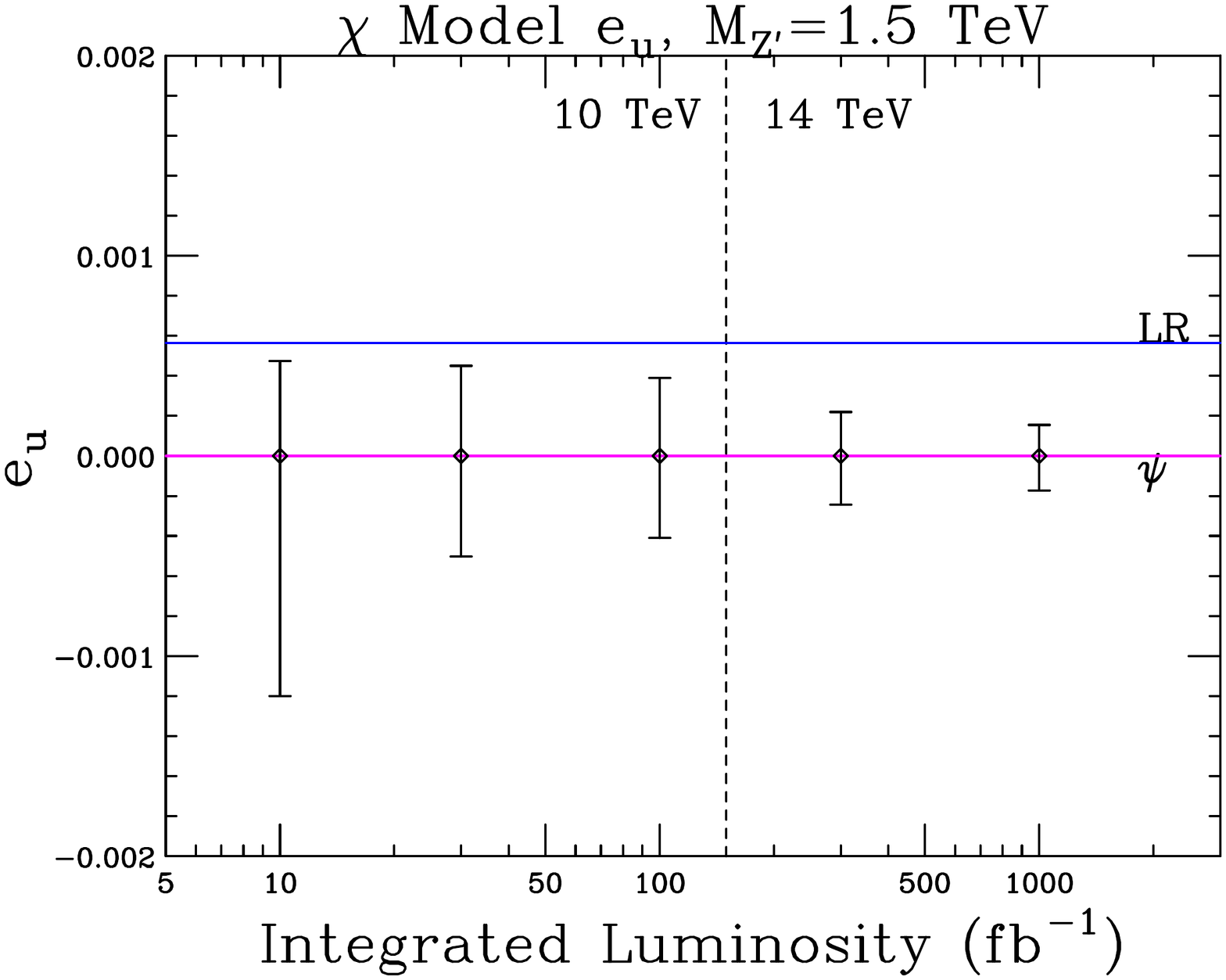}
\includegraphics[scale=0.45,angle=90]{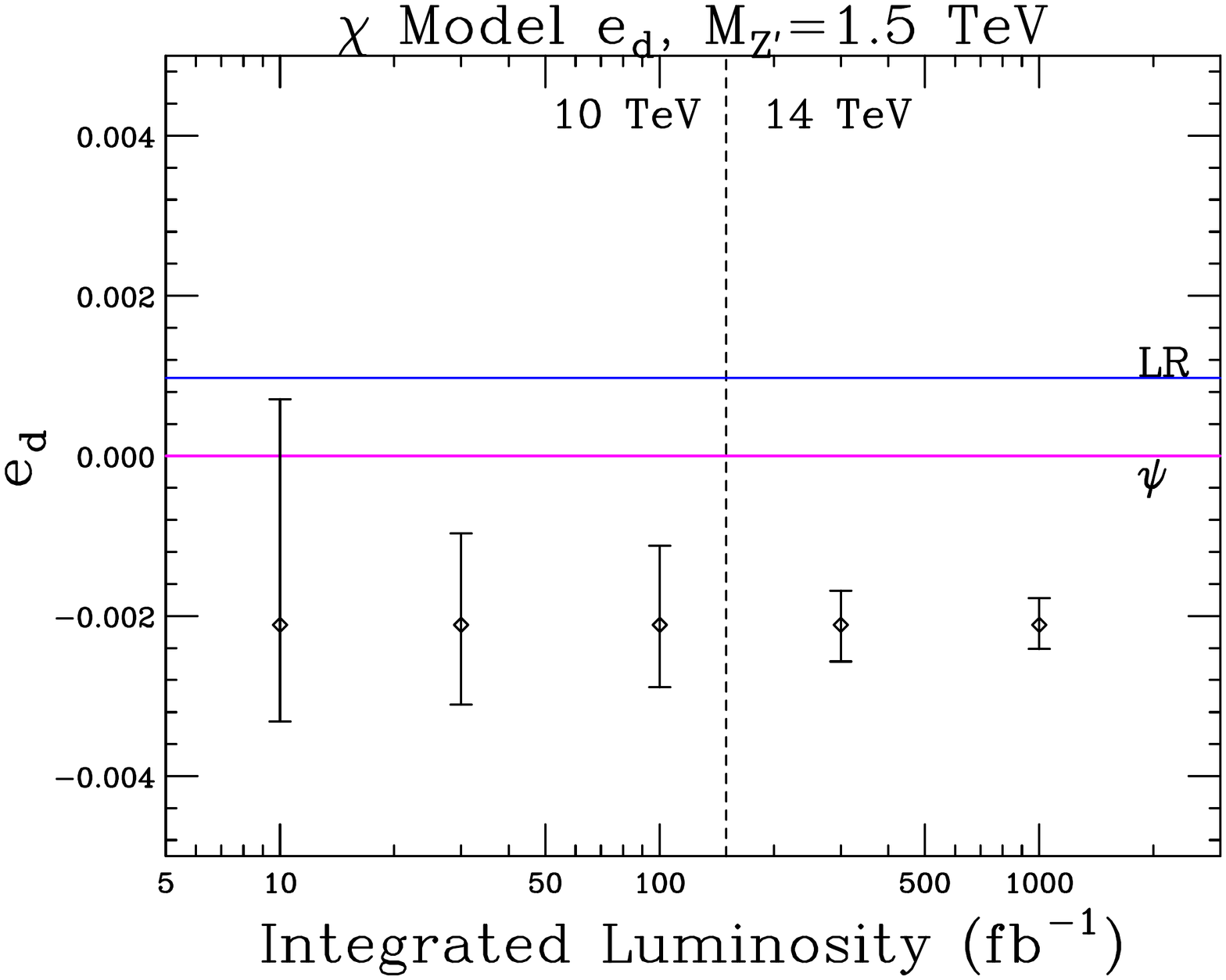} 
\caption{1$\sigma$ errors for the on-peak coupling combinations $e_u$ and $e_d$ as a function of integrated luminosity.  To the left of the dashed line, $\sqrt{s}=10$ TeV, while $\sqrt{s}=14$ TeV 
to its right.\label{e_1d}}
\end{figure}

\begin{figure}[htbp]
 \centering
\includegraphics[scale=0.45,angle=90]{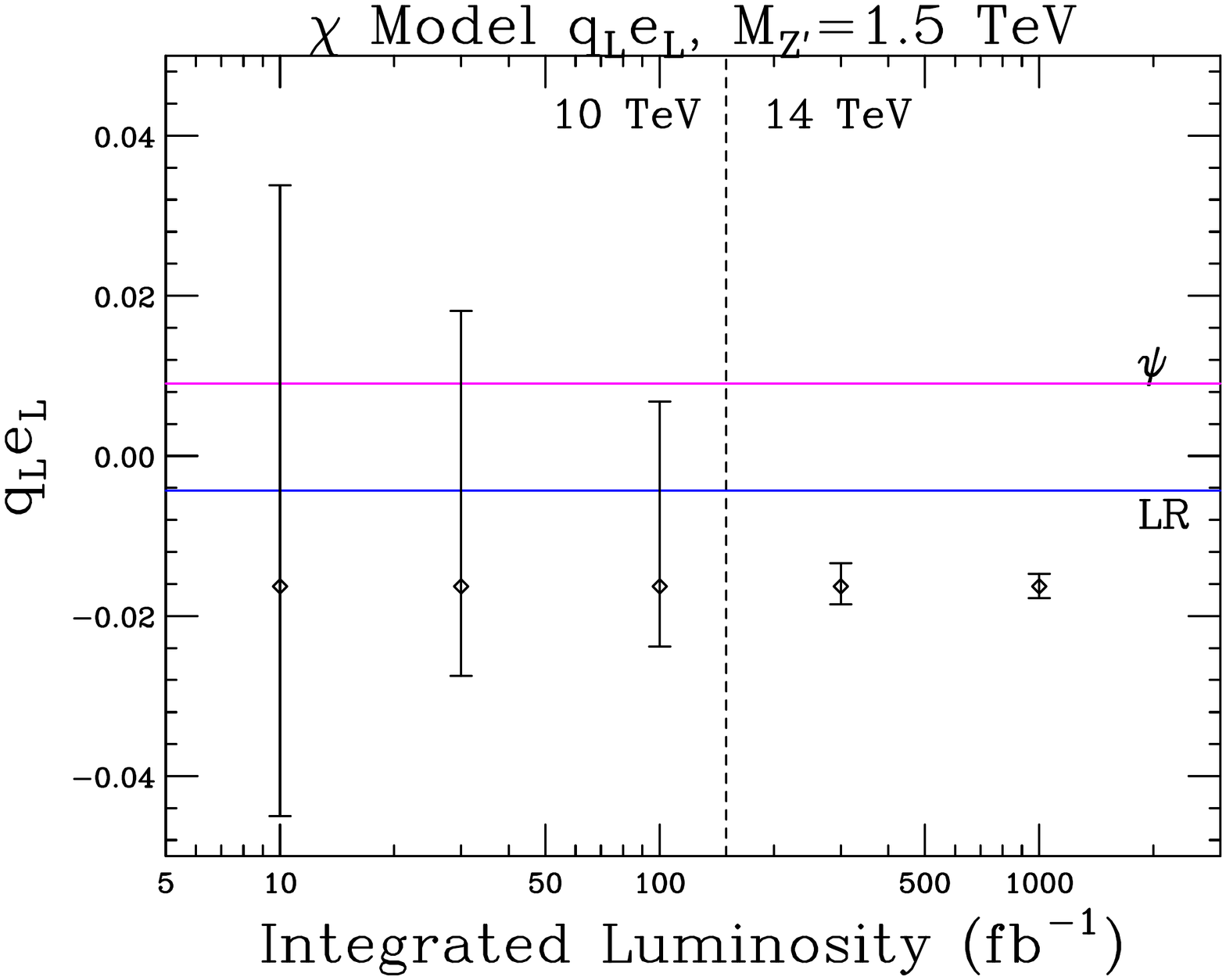}
\includegraphics[scale=0.45,angle=90]{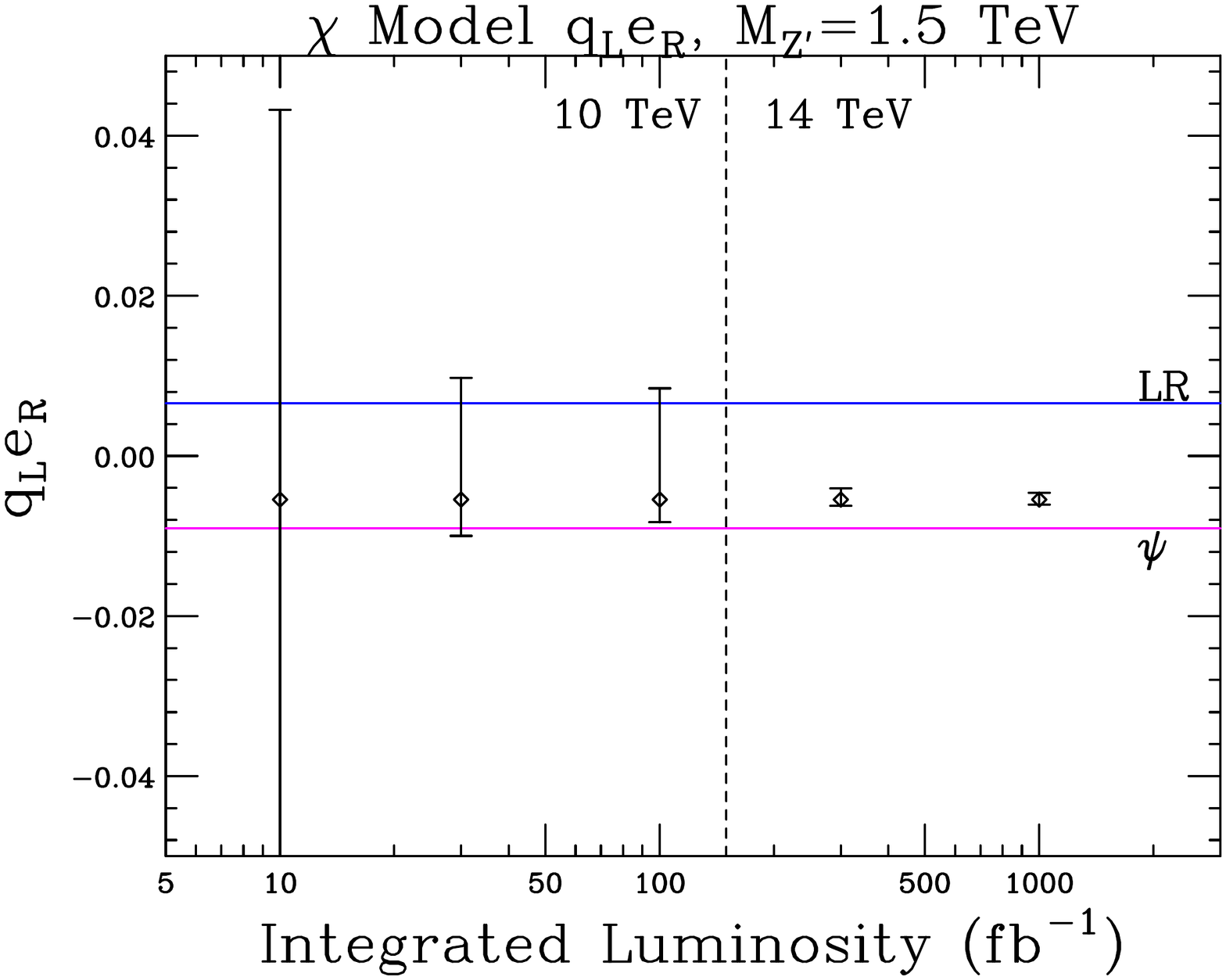} 
\caption{1$\sigma$ errors for the coupling combinations $q_Le_L,q_Le_R$ as a function of integrated luminosity.  Note that all signs are selected at 300 fb$^{-1}$, but not before.  To the left of the dashed line, $\sqrt{s}=10$ TeV, while $\sqrt{s}=14$ TeV 
to its right.\label{q_1d}}
\end{figure}

\begin{figure}[htbp]
 \centering
\includegraphics[scale=0.45,angle=90]{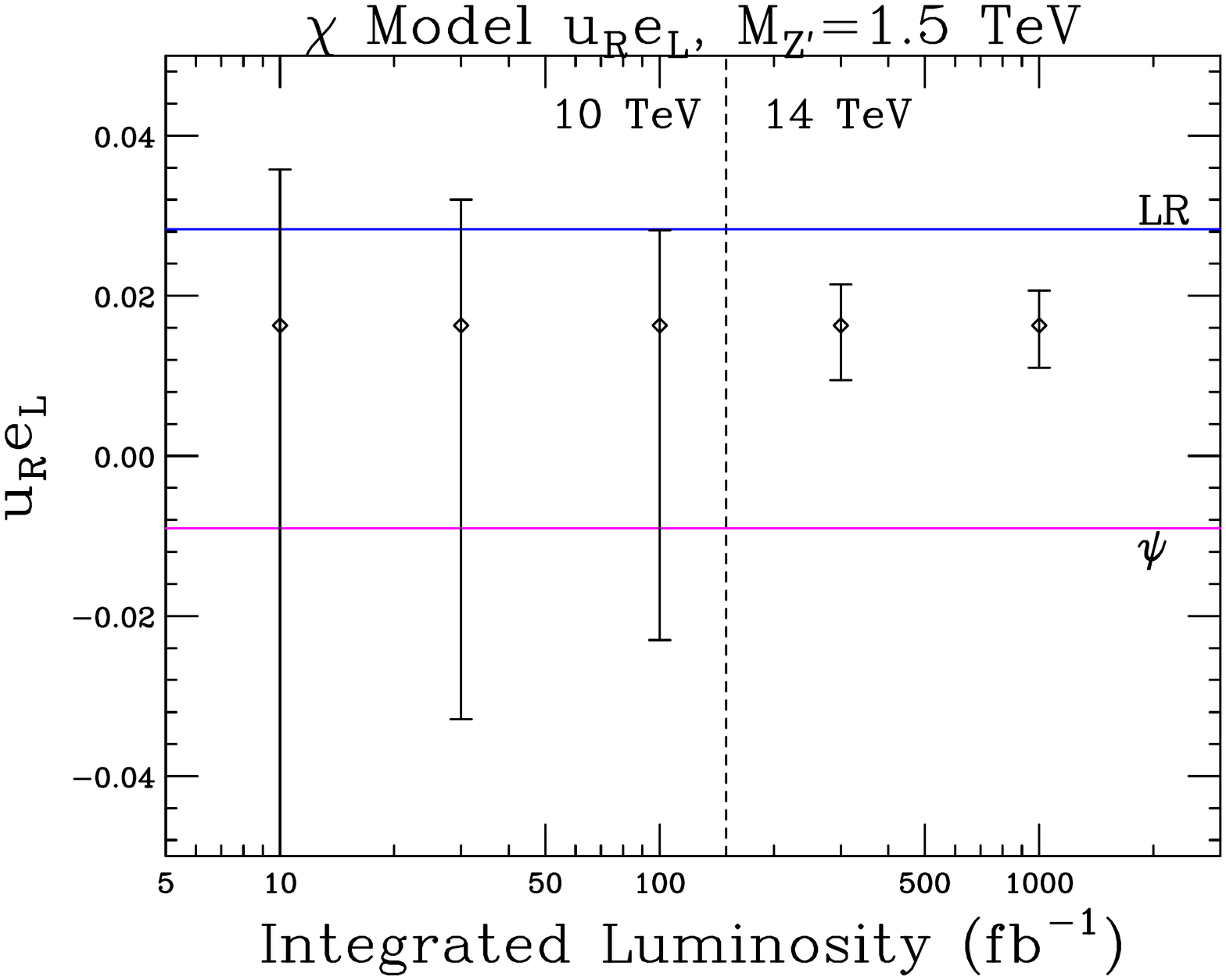}
\includegraphics[scale=0.45,angle=90]{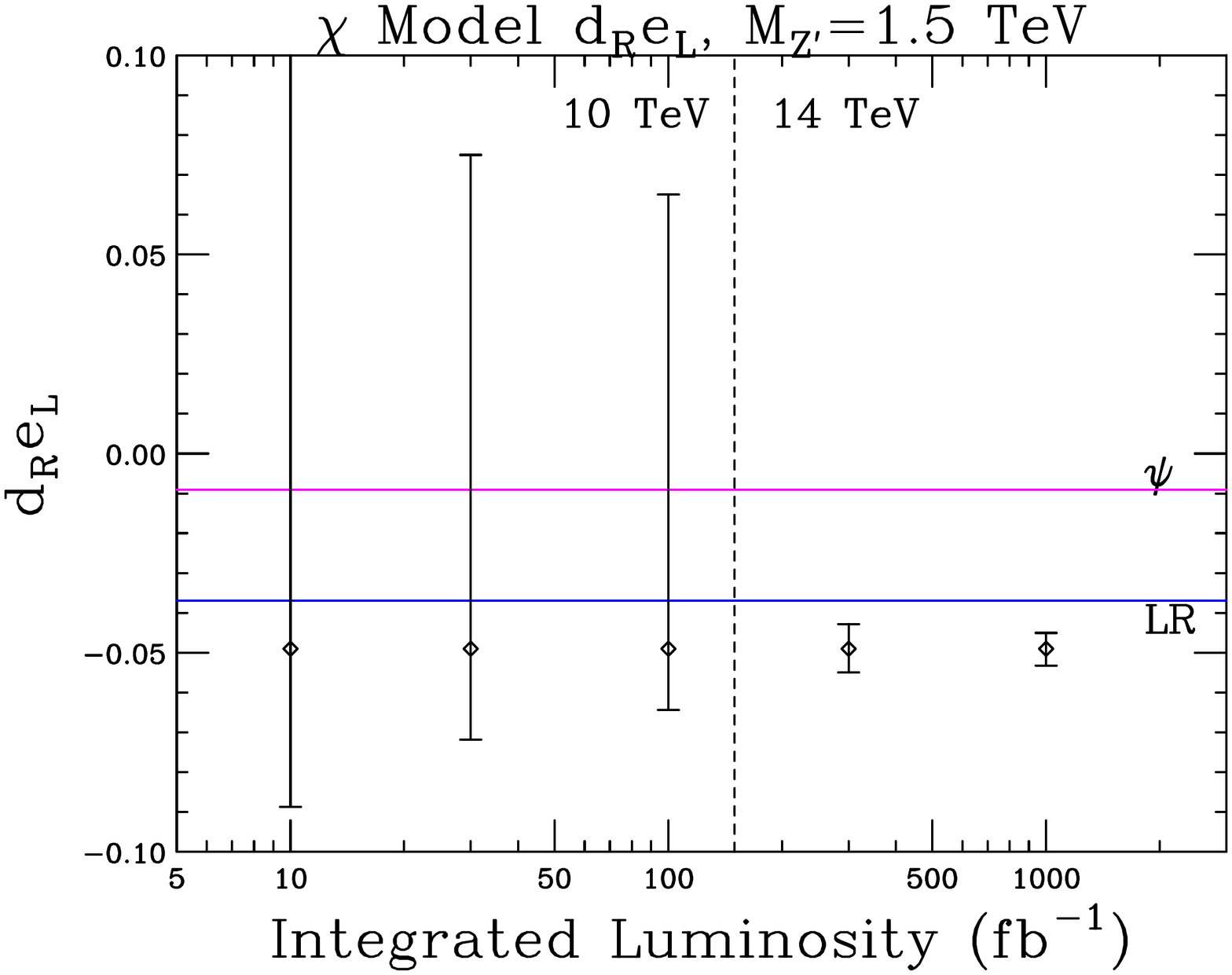} 
\caption{1$\sigma$ errors for the coupling combinations $u_Re_L,d_Re_L$ as a function of integrated luminosity.  Note that all signs are selected at 300 fb$^{-1}$, but not before.  To the left of the dashed line, $\sqrt{s}=10$ TeV, while $\sqrt{s}=14$ TeV 
to its right.\label{ud_1d}}
\end{figure}

We summarize the analyzing power of LHC and Q-weak by displaying in Figs.~\ref{c_1d},~\ref{e_1d},~\ref{q_1d},~and~\ref{ud_1d} the improvement in the coupling extraction for the $\chi$ model, one parameter at a time, as the luminosity is increased.  For integrated luminosities of 100 fb$^{-1}$ or less, we conservatively use a center-of-mass energy of 10 TeV; for higher luminosities we assume the full 14 TeV has been reached.  For the $c_q,e_q$ couplings, relative uncertainties of roughly $\pm 50\%$ for non-zero couplings are possible with 10 ${\rm fb}^{-1}$, while $\pm 30\%$ errors become possible with 30 ${\rm fb}^{-1}$.  At 300 ${\rm fb}^{-1}$ and above, the $q \times e$ combinations are directly measurable with precisions of $10\%$ or better.  We note that a sign degeneracy in all $q \times e$ parameters remains at integrated luminosities of less than 100 fb$^{-1}$; this is resolved at 1 $\sigma$ for all parameters at 300 fb$^{-1}$.

\section{Determining Leptonic Couplings}
\label{sec:moller}

Our goal is to probe all $Z'$ parameters to fully determine the underlying theory; thus far, we have only determined the couplings $q \times e$.  A linear collider can probe the leptonic couplings directly, even below the resonance peak, if the mass is known~\cite{Riemann:1996fk}.  In this paper, we do not assume access to a linear collider, and attempt to determine how well the last parameter can be measured by other means.  It is possible that observation of the $Z'$ decaying to bottom or top quark final states can help break this 
degeneracy.  We pursue here instead the alternate possibilities of determining the invisible width of the $Z'$, or by using a proposed low energy M\o{}ller scattering experiment.  The M\o{}ller scattering 
analysis in particular demonstrates a nice complimentarity between low and high-energy experiments.  A study of the sensitivity of 
future low energy M\o{}ller scattering experiments to $Z'$ effects has also been considered in Ref.~\cite{Chang:2009yw}, although the analysis performed there did not focus on the interplay with 
LHC measurements.

\subsection{Using the width \label{widthsec}}
If we assume that the leptonic couplings are flavor-universal, we can write the $Z'$ width in the limit of vanishing SM fermion masses in terms of parameters already determined, and the leptonic coupling:
\begin{align}
\Gamma_{Z'} & = \frac{M_{Z'}}{8 \pi} (6 q_L^2 + 3 u_R^2 + 3 d_R^2 + 2 e_L^2 + e_R^2) + \Gamma_{new} \notag \\
& = \frac{M_{Z'}}{8 \pi e_L^2} (6 (q_Le_L)^2 + 3 (u_Re_L)^2 + 3 (d_Re_L)^2 + (2+\tan^2\theta_l) e_L^4) + \Gamma_{new} .
\end{align}
We have denoted the decay width of the $Z'$ into particles other than SM fermions as $\Gamma_{new}$, even if those decays are into SM modes such as $W^+W^-$.  If this extra width is determine by measuring additional visible and invisible decays, we can extract $e_L^2$ up to a 2-fold degeneracy, with no further input\footnotemark:
\be
e_L^2 = \frac{4 \pi \frac{\Gamma_{Z'} - \Gamma_{new}}{M_{Z'}} \pm \sqrt{(4 \pi \frac{\Gamma_{Z'} - \Gamma_{new}}{M_{Z'}})^2 - (2+\tan^2\theta_l)(6 (q_Le_L)^2 + 3 (u_Re_L)^2 + 3 (d_Re_L)^2)}}{2+\tan^2\theta_l} .
\ee
The invisible width can be probed in associated production with a $Z$~\cite{Petriello:2008pu} or photon~\cite{Gershtein:2008bf} by comparing the branching fraction of invisible $Z'$ decays to leptonic $Z'$ decays, where the invisible decays due to a $Z'$ have been identified with a missing $p_T$ cut.  This would allow us to determine $\Gamma_{new}$.  The quality of this probe is highly model-dependent, so we do not include it in the analysis, but point out that it could be used to determine the leptonic couplings.

\footnotetext{If $e_L \ra 0$, one should instead solve for $e_R$.}

If the invisible width cannot be determined to any degree of accuracy, then in the worst case we have bounds for $e_L^2$.  For a given set of coupling combinations $q \times e$, raising (or lowering) $e$ will raise the partial width into $e$ (or $q$).  Since we have measured the width, there is a natural ceiling in either direction, corresponding to the bounds

\begin{align}
e_L^2 \geq \frac{4 \pi \frac{\Gamma_{Z'}}{M_{Z'}} - \sqrt{(4 \pi \frac{\Gamma_{Z'}}{M_{Z'}})^2 - (2+\tan^2\theta_l)(6 (q_Le_L)^2 + 3 (u_Re_L)^2 + 3 (d_Re_L)^2)}}{2+\tan^2\theta_l}, \\
e_L^2 \leq \frac{4 \pi \frac{\Gamma_{Z'}}{M_{Z'}} + \sqrt{(4 \pi \frac{\Gamma_{Z'}}{M_{Z'}})^2 - (2+\tan^2\theta_l)(6 (q_Le_L)^2 + 3 (u_Re_L)^2 + 3 (d_Re_L)^2)}}{2+\tan^2\theta_l}.
\end{align}
These bounds are also highly model-dependent, as they are much looser if there is a lot of ``extra'' width.  This is easily achieved in models with additional exotic particles into 
which the $Z'$ can decay~\cite{Arvanitaki:2006cy}.  As such, we refrain from putting these bounds in our plots, though they are always present.  We also note that 
if the $Z'$ width is determined, upper limits on all couplings can be derived by requiring that a given partial width no more than saturates the measured value.

\subsection{M\o{}ller scattering}
The future J-Lab M\o{}ller scattering experiment may be able to observe a shift in the asymmetry as discussed previously.  Measuring the combination $e_V e_A$ amounts to measuring a hyperbola in the $e_L - e_R$ plane: $e_V e_A = (e_R^2 - e_L^2)/4$.  Using the anticipated experimental and theoretical errors,
\be
\delta \frac{e_R^2 - e_L^2}{M_{Z'}^2} = 0.022 \rm{ TeV}^{-2},
\ee
or, for a 1.5 TeV $Z'$,
\be
\delta (e_R^2 - e_L^2) = 0.050 .
\ee
The shifts due to our considered models are listed below.  As with Q-weak, we ignore running down from the $Z'$ scale.

\begin{table}[ht]
 \begin{center}
  \begin{tabular}{| c | c |}
\hline
& $e_R^2 - e_L^2$ \\ \hline
$\chi$ & -0.0436 \\ \hline
$LR$ & 0.0167 \\ \hline
$\psi$ & 0 \\ \hline
$\chi^*$ & -0.0833 \\ \hline
  \end{tabular}

 \end{center}
\caption{Coupling combinations contributing to M\o{}ller scattering for our various test cases.}
\end{table}

\begin{figure}[htbp]
 \centering
\includegraphics[scale=0.6,angle=90]{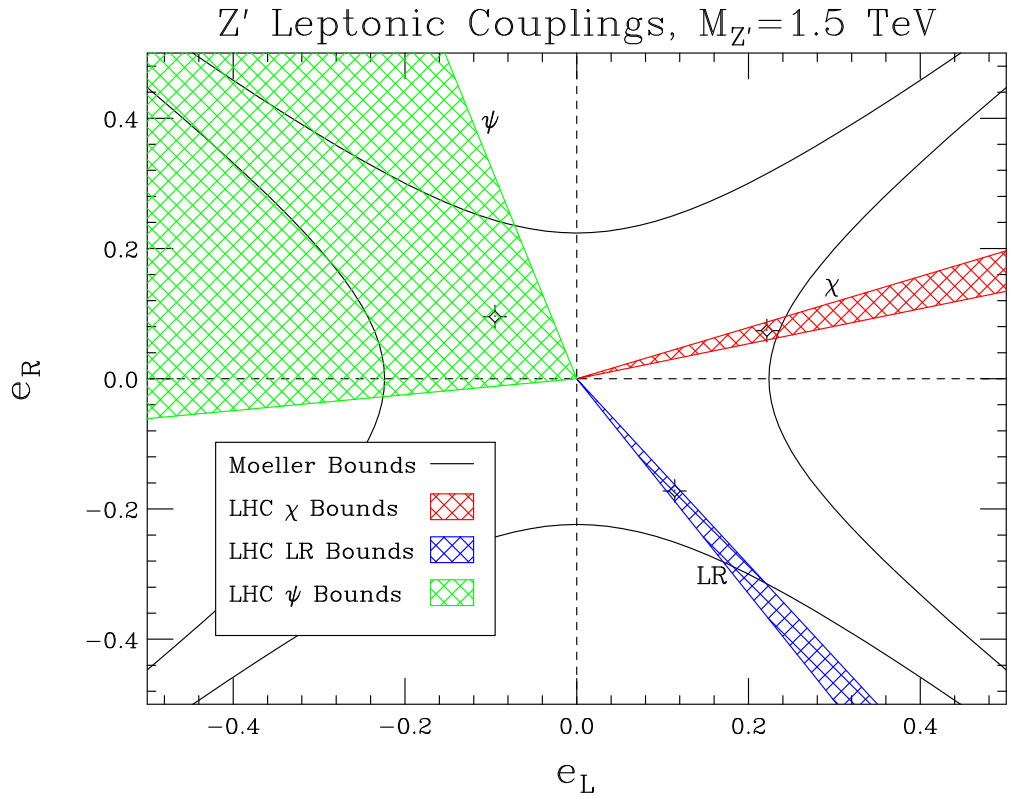}
\caption{Measurements of the leptonic couplings from the LHC and J-Lab M\o{}ller experiment.  M\o{}ller bounds assume no measured deviation from the Standard Model.  The overall, unphysical, sign of the coupling is chosen for each model for convenient display of the results.}
\label{mollerplot}
\end{figure}

None of the test models with the standard coupling induces a $1 \sigma$ deviation in M\o{}ller.  In Fig.~\ref{mollerplot}, we plot these models with the $1 \sigma$, 1 ab$^{-1}$ LHC errors in $\theta_l$, as well as hyperbolic upper limits on the size of the leptonic couplings assuming no deviation from the Standard Model is found.  The allowed regions do not actually extend to the origin; there is some mininum size of the leptonic couplings due to the argument of Sec.~\ref{widthsec}.  An overall sign in the couplings is unphysical, so we have chosen the sign of $e_L$ for each model for convenience in displaying the plot.

\begin{figure}[htbp]
 \centering
\includegraphics[scale=0.6,angle=90]{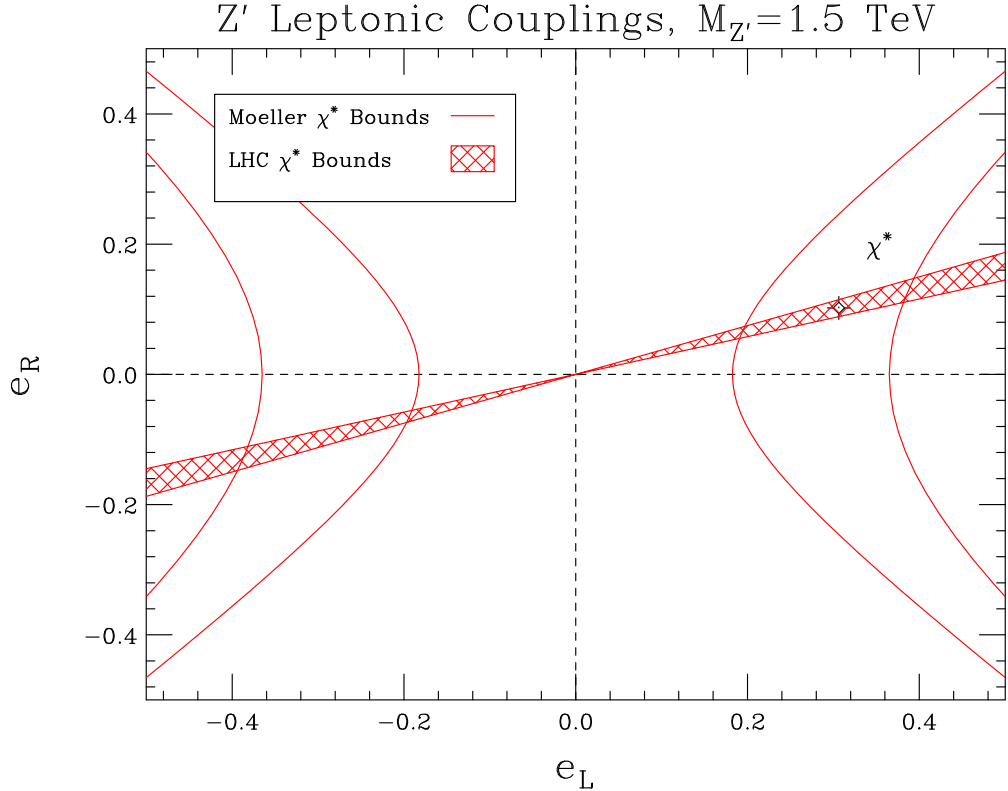}
\caption{Measurements of the leptonic couplings from the LHC and J-Lab M\o{}ller experiment, for a model with larger couplings.  There are upper and lower hyperbolic bounds on the size of the couplings.}
\label{mollerhalfplot}
\end{figure}

Our original test models should be consistent with the Standard Model in the M\o{}ller experiment for a mass of 1.5 TeV at $1 \sigma$, though the $\chi$ model is very close to the limit.  We therefore ask the question, if the leptonic couplings were somewhat larger, what could we see?  As an example, we consider a test model like the previously considered $\chi$ model, except with a larger coupling; we call this the $\chi^*$ model.  Specifically, we replace the $e/\cos \theta_W$ appearing in $E_6$ models with a somewhat larger value, 1/2.  Such shifts in the overall normalization from that predicted by the canonical breaking pattern has been considered elsewhere in the literature~\cite{Strassler:2006im}.  In this case, the leptonic asymmetry is large enough to produce a significant deviation from the standard model in the M\o{}ller experiment.  This produces upper and lower hyperbolic bounds in the $e_L$ - $e_R$ plane.  We plot the expected results in Fig.~\ref{mollerhalfplot}.   
M\o{}ller asymmetry gives us a hyperbola in the $e_L$ - $e_R$ plane.  The LHC gives us the angle.  Together, they pinpoint the $Z'$ leptonic couplings to a small region in the $e_L$ versus $e_R$ plane.

\subsection{Putting it all together}

Once the leptonic couplings are determined, they can be input into the $q \times e$ measurement coming from the LHC and Q-weak to extract all $Z'$ couplings.  Explicitly, the quark charges are given by
\begin{eqnarray}
q_L &= & \frac{q_L e_L}{e_L}, \nonumber \\
u_R &= & \frac{u_R e_L}{e_L}, \nonumber \\
d_R &= & \frac{d_R e_L}{e_L} .
\end{eqnarray}
We list in Table~\ref{fintab} the final expected results after solving these equations for the quark charges and propagating through the anticipated M\o{}ller errors.  We include all LHC data at 1 ab$^{-1}$, and the Q-weak and M\o{}ller experiments.   Each parameter range given is 68\% CL.  Note that for typical models, we only get upper bounds on the sizes of the leptonic couplings, and lower bounds for the sizes of the quark couplings.  The signs are determined from the LHC.  We note that the bounds derivable from width constraints discussed above have not been included in this table.

\begin{table}[ht]
 \begin{center}
  \begin{tabular}{| c | c | c | c | c |} \hline
   & $\chi$ & LR & $\psi$ & $\chi^*$ \\ \hline
$e_L$ & 0 to $0.323$ & 0 to $0.228$ & + & $0.306 \pm^{0.074}_{0.100}$ \\ \hline
$e_R$ & 0 to $0.111$ & $-0.343$ to 0 & - & $0.102 \pm^{0.028}_{0.034}$ \\ \hline
$q_L$ & $< -0.051$ & $< -0.009$ & + & $-0.102 \pm^{0.020}_{0.051}$ \\ \hline
$u_R$ & $> 0.045$ & $> 0.126$ & - & $0.102 \pm^{0.057}_{0.029}$ \\ \hline
$d_R$ & $< -0.150$ & $< 0.162$ & ? & $-0.306 \pm^{0.150}_{0.061}$ \\ \hline
  \end{tabular}
 \end{center}
\caption{Final determinations of the couplings using data from the LHC, Q-weak, and low energy M\o{}ller scattering.  The results assume 1 ab$^{-1}$ of integrated luminosity at the LHC.  The sign of $e_L$ is chosen to be positive to fix the overall sign degeneracy.  $d_R$ remains totally unknown for the $\psi$ model extraction, and for other parameters only the sign is determined. \label{fintab}}
\end{table}

\section{Conclusions}
\label{sec:conc}

We have presented a thorough look at the ability of the LHC to analyze the underlying model parameters of a new $Z^{\prime}$ boson in the dilepton channel.  In addition, we have studied 
the complimentary information that can be obtained from upcoming high-precision, low-energy experiments.  Our analysis is as model-independent as possible, and includes the major sources of error 
which hinder the parameter extraction.  We have discussed the important role played by off-peak data and the future Q-weak measurement of the proton weak charge in removing sign degeneracies that 
remain if only LHC data on the $Z^{\prime}$ peak is analyzed.  We also have shown how a precision measurement of low-energy M\o{}ller scattering can break the $q \to yq, e\to e/y$ 
scaling degeneracy inherent to the dilepton channel at the LHC, and permit an extraction of the individual $Z^{\prime}$ charges.  Our study demonstrates that a global analysis of data from 
many different observations is a necessary component of any attempt to determine $Z'$ couplings from the bottom up.

Our study has elucidated the how precisely $Z^{\prime}$ couplings can be measured as a function of integrated luminosity.  Our LHC analysis relies upon appropriate binning of dilepton events 
in rapidity and scattering angle: $u$ and $d$ couplings are correlated with higher and lower $Z'$ rapidities respectively, while left- and right-handed couplings yield different lepton angular distributions.  As the parameters extracted and the bins used vary as a function of integrated LHC lumunosity, we summarize below how we believe the analysis of a dilepton 
signal at the LHC should proceed.  This template assumes that the resonance has a mass, width and total cross section roughly in the ranges considered here.  For significantly different values, the 
timeline presented below must be adjusted.
\begin{itemize}

\item With less than 10 ${\rm fb}^{-1}$, the first property to establish is the spin of the observed resonance~\cite{Allanach:2000nr,Osland:2009tn}.  At this point, a $\chi^2$ comparison between 
data and model hypotheses based on the three on-peak observables discussed in Section~\ref{sec:chi2} allows initial guesses as to the underlying $Z^{\prime}$ model to be tested.  This analysis is 
less sensitive to additional decay modes and overall coupling than comparisons of the total cross sections.

\item With $10-30$ ${\rm fb}^{-1}$, enough on-peak data is available to attempt an extraction of the couplings.  Sufficient data exists only on-peak at this point, indicating that the natural parameters 
to study are the $c_q$, $e_q$ defined in Eq.~(\ref{cedef}).  This analysis can be performed in a simple, model-independent fashion by following the procedure described in Section~\ref{sec:initlhc}.  We have found that the $c_q$ coupling combinations can be measured with relative uncertainties of $\pm 30\%$ for the $M_{Z^{\prime}}=1.5$ TeV test cases considered.

\item On-peak observables are sensitive to only the combinations $(q \times e)^2$.  The sign ambiguities can be removed by studying off-peak data, which probes the $q \times e$ combinations 
directly.  This becomes possible with $100$ ${\rm fb}^{-1}$ of integrated luminosity.  We have found that $q \times e$ combinations can be determined with relative uncertainties of approximately 10 to 100\%, depending on the particular model and coupling combination, for luminosities of $300-1000$ ${\rm fb}^{-1}$.  The couplings display very strong correlations, which indicate a much stronger reduction in parameter space than the individual errors would suggest.  A precision measurement from Q-weak helps remove sign degeneracies that remain after LHC running.  Another interesting aspect we have seen in our study is that the width can be determined precisely at this stage without a direct measurement of the Breit-Wigner shape. 

\item Some direct measurement of either lepton or quark couplings is crucial to break the $q \to yq, e\to e/y$ degeneracy that exists in the LHC dilepton mode.  We have found that a proposed M\o{}ller experiment can determine the combination $e_R^2 - e_L^2$ directly in some situations, and bound it in others.  Measuring this last parameter can tell us all the $Z'$ couplings to Standard Model fermions, by separating out the well-measured combinations $q \times e$ from the LHC.

\end{itemize}

Our study can be extended in several interesting ways.  It would be interesting to apply our analysis procedure to other $Z^{\prime}$ models to more fully determine its robustness.  Another possible extension would be to add other LHC final states such as $t\bar{t}$, which could also potentially break the scaling degeneracy from which the dilepton channel suffers.  We look forward to studying these possibilities and hope that Nature provides on opportunity to apply our procedure to LHC data.

\bigskip
\bigskip
\noindent
{\bf{\Large Acknowledgements}}
\bigskip

\noindent
We thank J. Erler, M. Herndon, K. Kumar, M. Ramsey-Musolf and T. Rizzo for many helpful discussions and suggestions.  This work is supported by the DOE grant DE-FG02-95ER40896, by the University of Wisconsin Research Committee with funds provided by the Wisconsin Alumni Research Foundation, and
by the Alfred P.~Sloan Foundation.

\end{document}